\documentclass[sigconf]{acmart}

\usepackage{amsmath,amsfonts}
\usepackage{algorithm}
\usepackage{adjustbox}
\usepackage{lscape}
\usepackage{siunitx}
\usepackage[noend]{algpseudocode}
\usepackage{textcomp}
\usepackage{fancyvrb}
\usepackage{graphicx}
\usepackage{soul}
\usepackage{booktabs}
\usepackage[utf8]{inputenc}
\usepackage[T1]{fontenc}
\usepackage{rotating}
\usepackage{graphicx}
\usepackage{paralist}
\usepackage{tabularx}
\usepackage{multicol}
\usepackage{multirow}
\usepackage{pbox}
\usepackage{enumitem}	
\usepackage{colortbl}
\usepackage{pifont}
\usepackage{xspace}
\usepackage{url}
\usepackage{tikz}
\usepackage{tabularx}
\usepackage{fontawesome}
\usepackage{lscape}
\usepackage{listings}
\usepackage{color}
\usepackage{showexpl}
\usepackage{anyfontsize}
\usepackage{comment}
\usepackage{soul}
\usepackage{multibib}
\usepackage{multirow}
\usepackage{xcolor}
\usepackage{xspace}
\usepackage{caption}
\usepackage{footnote}
\usepackage{tcolorbox}
\usepackage{booktabs}
\usepackage{fixltx2e}
\usepackage{balance}
\usepackage[normalem]{ulem}

\AtBeginDocument{%
  }

\newboolean{showcomments}

\setboolean{showcomments}{true}

\ifthenelse{\boolean{showcomments}}
{\newcommand{\nb}[2]{
		\fbox{\bfseries\sffamily\scriptsize#1}
		{\sf\small$\blacktriangleright$\textit{#2}$\blacktriangleleft$}
	}
	
}
{\newcommand{\nb}[2]{}
	
}

\newcommand{\rev}[1]{\textcolor{black}{#1}}

\newcommand{\tool}{{GH-WCOM}\xspace}

\newcommand{\ie}{\emph{i.e.,}\xspace}
\newcommand{\eg}{\emph{e.g.,}\xspace}
\newcommand{\etc}{etc.\xspace}
\newcommand{\etal}{\emph{et~al.}\xspace}
\newcommand{\secref}[1]{Section~\ref{#1}\xspace}
\newcommand{\figref}[1]{Fig.~\ref{#1}\xspace}

\newcommand{\tabref}[1]{Table~\ref{#1}\xspace}

\newcommand{\java}{\emph{Java}\xspace}
\newcommand{\jc}{\emph{$JC_{task}$ }\xspace}
\newcommand{\cl}{\emph{$NS_{task}$ }\xspace}
\newcommand{\tfivemodel}[1]{T5$_{\mathit{#1}}$\xspace}
\newcommand*\circled[1]{\tikz[baseline=(char.base)]{
		\node[shape=circle,fill,inner sep=0.8pt] (char) {\textcolor{white}{#1}};}}

\definecolor{lightergray}{rgb}{0.9,0.9,0.9}
\newtcolorbox{resultbox}{colback=lightergray, arc=0.5mm, top=2mm, bottom=2mm, left=2mm, right=2mm}

\begin{document}

\title{Toward Automatically Completing GitHub Workflows}


\author{Antonio Mastropaolo}
\email{antonio.mastropaolo@usi.ch}
\affiliation{%
  \institution{SEART @ Software Institute, \\Universit\`a della Svizzera Italiana}
  \city{Lugano}
  \state{Switzerland}
  \country{CH}
}

\author{Fiorella Zampettti}
\email{f.zampetti@unisannio.it}
\affiliation{%
	\institution{Dept. of Engineering, \\University of Sannio}
	\city{Benevento}
	\state{Italy}
	\country{IT}
}

\author{Gabriele Bavota}
\email{gabriele.bavota@usi.ch}
\affiliation{%
	\institution{SEART @ Software Institute, \\Universit\`a della Svizzera Italiana}
	\city{Lugano}
	\state{Switzerland}
	\country{CH}
}

\author{Massimiliano Di Penta}
\email{dipenta@unisannio.it}
\affiliation{%
	\institution{Dept. of Engineering, \\University of Sannio}
	\city{Benevento}
	\state{Italy}
	\country{IT}
}

\begin{abstract}
Continuous integration and delivery (CI/CD) are nowadays at the core of software development. Their benefits come at the cost of setting up and maintaining the CI/CD pipeline, which requires knowledge and skills often orthogonal to those entailed in other software-related tasks. While several recommender systems have been proposed to support developers across a variety of tasks, little automated support is available when it comes to setting up and maintaining CI/CD pipelines. We present \tool (GitHub Workflow COMpletion), a Transformer-based approach supporting developers in writing a specific type of CI/CD pipelines, namely GitHub workflows.To deal with such a task, we designed an abstraction process to help the learning of the transformer while still making \tool able to recommend very peculiar workflow elements such as tool options and scripting elements. Our empirical study shows that \tool provides up to 34.23\% correct predictions, and the model's confidence is a reliable proxy for the recommendations' correctness likelihood.  
  
\end{abstract}



\maketitle


\section{Introduction} \label{sec:intro}

Setting and maintaining a continuous integration and delivery (CI/CD) pipeline is crucial for any software project. Indeed, CI/CD contributes to enhancing software quality and developers' productivity~\cite{CHEN201772}, and to speed up release cycles~\cite{VasilescuFSE15}. Nevertheless, previous research has highlighted the challenges encountered by developers in setting up and maintaining CI/CD pipelines~\cite{Chen:2015,HiltonFSE17,zampetti2020empirical,zampetti2022problems,saroar2023developers}.  
Despite the availability of modern CI/CD infrastructures and reusable assets (\eg GitHub actions), the intrinsic CI/CD requirements and underlying technology of a given project may still make this task hard~\cite{HiltonFSE17,zampetti2022problems}. For example, this could be the case when a system needs to be deployed and tested on different operating systems or even embedded devices. 

The aforementioned challenges entail the need for recommender systems helping developers in setting up and maintaining CI/CD pipelines. \rev{This is also supported by a study by Soroar \etal \cite{saroar2023developers}, reporting that $\sim$60\% of the 90 developers they surveyed encountered difficulties in automating workflows using GitHub actions.}  

It is worth mentioning that the possible solutions are somewhat similar to those related to automated code completion, where approaches have been defined either to provide suggestions about non-trivial, generic code elements (up to blocks) to be completed \cite{ciniselli2021empirical}, or more specialized suggestions, \eg related to creating assertions \cite{WatsonTMBP20}, or repairing vulnerabilities \cite{ChenKM23,FuTLNP22} and bugs \cite{ChenKTPPM21,li2020repair,li2022repair}.


That being said, helping developers in setting up a CI/CD pipeline poses unique challenges. Indeed, the structure a CI/CD pipeline mixes up very specific scripting elements (\eg related to configuring a server, downloading certain libraries, \etc) with some more recurring and regular reusable elements (\eg the actions in the case of GitHub), up to natural language elements. Also, CI/CD pipeline contain several extremely context-specific elements, such as paths of installation directories, or URLs of resources to download.
This creates major challenges to the use of data-driven techniques for the automated recommendations of these elements.

This paper proposes \tool (GitHub Workflow COMpletion) an approach leveraging Transformer models \cite{vaswani2017attention} to provide automated completion of GitHub workflows. To develop (and train) \tool, we have leveraged the existing body of GitHub workflows starting from a dataset by Decan \etal~\cite{DecanMMG22}. 

\eject

To make a GitHub workflow completion possible, we have defined and implemented a multi-step pre-processing including an abstraction of the tokens for which their verbatim prediction would not be feasible (\eg a very specific path in a project) while still leaving to \tool the ability to recommend some very peculiar workflow elements such as tool options and other scripting elements. \tool can recommend GitHub workflow completions in different modes that mimic how a developer may implement the workflow, \ie (i) suggesting the next statement (a GitHub step), or (ii) helping to complete a job with implementation elements once the developer has defined, in plain English, what the job should do.

Summarizing, this paper makes the following contributions:

\begin{compactenum}
	\item We propose \tool, which, to the best of our knowledge, is the first approach to automatically complete CI/CD pipelines, and GitHub workflows in particular.
	\item We experiment with different pre-trainings, abstraction levels, and completion scenarios. Results indicate that pre-training at least on English text is required, and \tool's performance for correct prediction is $\sim$34\%. The correct prediction accuracy is correlated with the model's confidence.
	\item We report a qualitative analysis discussing the extent to which the recommendations provided by \tool could still be helpful also when the generated output is different from the target (expected) one. Also, we discuss how \tool is competitive with respect to recent, popular general-purpose recommenders based on large language models, \eg CoPilot \cite{copilot} and ChatGPT \cite{chatgpt}.
	\item We made publicly available \tool scripts, checkpoints predictions, and the used datasets~\cite{replication}.
\end{compactenum}

\section{Background}
\label{sec:back}

\begin{figure}[t]
	\centering
	\includegraphics[width=0.65\columnwidth]{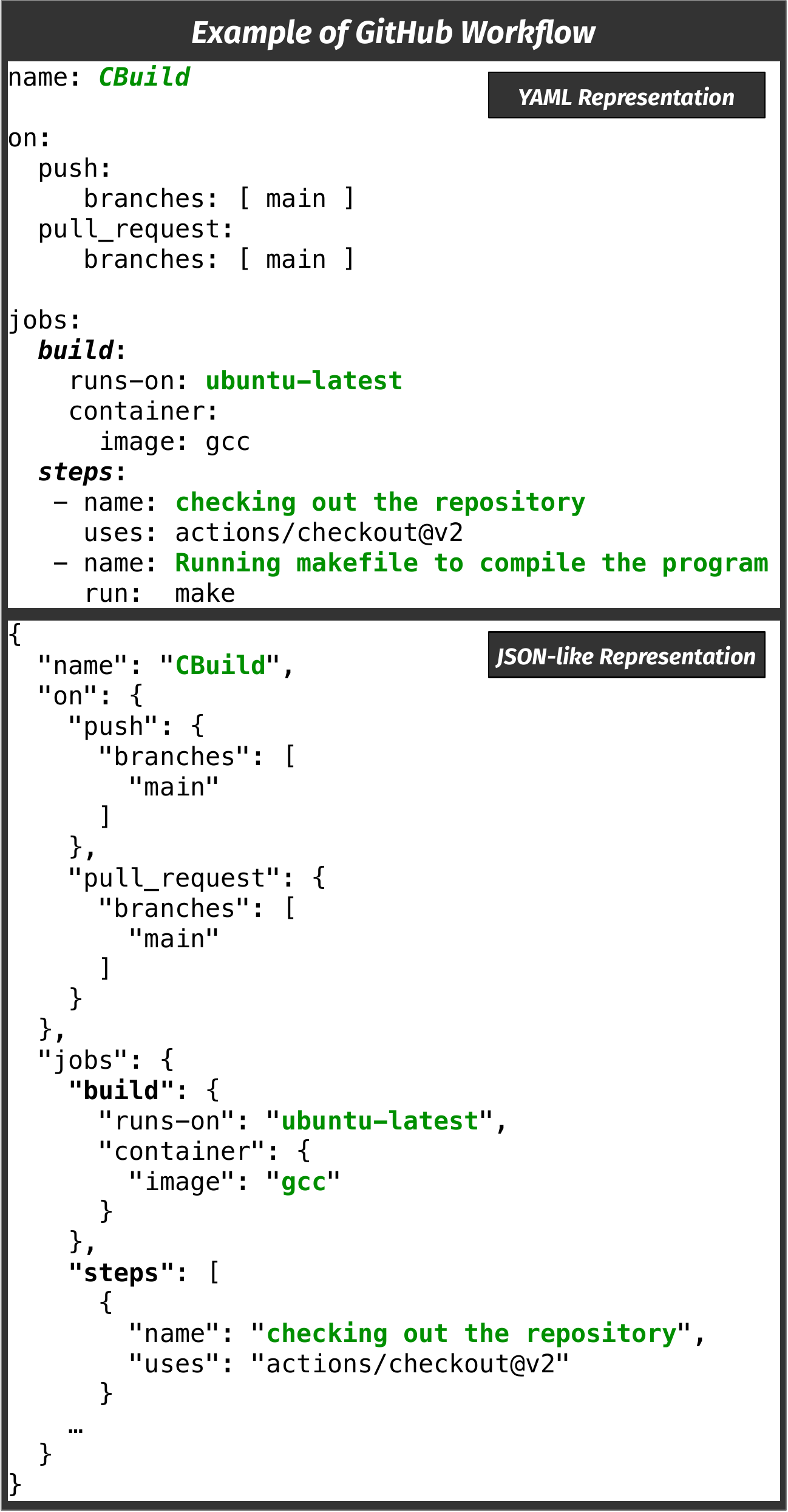}
	\vspace{-0.2cm}
	\caption{GitHub workflow example}
	\label{fig:workflow-example}
	\vspace{-0.1cm}
\end{figure}

GitHub workflows integrate CI/CD in the GitHub infrastructure.
A GitHub workflow (example in the top part of \figref{fig:workflow-example}, while the bottom part will be described later in the paper) is a YAML file located under the \texttt{.github/workflows} (sub)directory of a repository.
As specified by the \texttt{on:} clause, a workflow is triggered based on some events (\eg a push, a pull request) and executes a series of jobs, specified after the \texttt{jobs} keyword (as the job named \texttt{build} in the figure). 

Jobs are units of execution of a CI/CD process and can run in parallel or sequentially (if dependencies between jobs are specified) on runners. Unless they use explicit ways to exchange information (\eg uploading and downloading artifacts in a storage area), jobs are independent of each other. Runners can be local or remote virtual machines or containers.
Runners and containers are specified after the job name, using the \texttt{runs-on} clause, and, if containers are used, the \texttt{container:} and \texttt{image} clauses. The job in the example runs on an Ubuntu virtual machine and uses a container from an image bringing the \emph{gcc} compiler.
Each job consists of a sequence of steps. In \figref{fig:workflow-example}, steps are all items preceded by a dash following the keyword \texttt{steps}.
There are two ways to implement a step. The first (denoted by the keyword \texttt{uses}) is to leverage GitHub actions, \ie reusable applications available on GitHub that implement recurring tasks. For example, the \texttt{actions/checkout@v2} is version 2 of an action checking the content of the GitHub repository branch on which the workflow has been triggered.

The second (keyword \texttt{run}) consists of directly executing whatever application is available in the virtual machine/container (\eg \texttt{apt-get} to install components, \texttt{gradle}  to run a Gradle build). Run steps are typically used for specific tasks for which an action is not available, or the task is so simple as to not require an action. Optionally, a step can be documented with a textual description of its action or run command, using the \texttt{name} keyword. Further information about GitHub workflows and actions is available on the GitHub documentation \cite{github-workflows}.


\section{\tool} \label{sec:approach}
This section describes \tool, the proposed approach to recommend GitHub workflow completions. \tool leverages the  Text-to-Text Transfer Transformer (T5) model by Raffel \etal \cite{raffel2019exploring}. First, we pre-train T5 by experimenting with different strategies. Then, we train the tokenizer needed by \tool and, after an hyperparameter calibration, we fine-tune T5 with instances specifically related to the actual prediction tasks. After that, we use the trained model for two different kinds of predictions, \ie (i) adding the next step in a workflow job, or (ii) completing a job whose steps have just been specified in terms of natural language text.

In the following, after overviewing the T5 model, we describe the different steps of the approach.


\subsection{An overview of T5}\label{sub:t5-overview}
T5 \cite{raffel2019exploring} is an encoder-decoder Transformer \cite{vaswani2017attention} designed to work in a text-to-text setting.  Whatever the generation task is, T5 can be employed as long as both the input and the output can be represented as textual strings (\eg translating from English to Spanish, outputting the fixed version of a provided buggy code). We have chosen T5 given its successful application in several code completion/generation tasks \cite{mastropaolo2022using, ciniselli2021empirical, tufano2022using, wang2021codet5}.

The training procedure of T5 is usually performed in two steps. First, the model is pre-trained on a large-scale dataset using self-supervised training. The pre-training provides T5 with general knowledge about the language(s) of interest. For example, assuming the will of building an English-to-Spanish translator, we could provide as an input to the model English and Spanish sentences having 15\% of their tokens masked, with the model in charge of predicting them. That makes the pre-training fully self-supervised.

Subsequently, the model undergoes fine-tuning, which is supervised training (\eg providing pairs composed of an English sentence and its Spanish translation). Fine-tuning specializes the model for the task of interest. 
 
Raffel \etal experimented with five T5 variants, differing in terms of the number of trainable parameters: small, base, large, 3 billion, and 11 billion. Considering our computational resources and recent successful application of T5$_{small}$ to automate code-related tasks \cite{mastropaolo2022using, ciniselli2021empirical, tufano2022using, wang2021codet5}, we opted for the simplest architecture which still features 60M trainable parameters, consistently with large language models used in the literature. For additional architectural details, we point the reader to the work by Raffel \etal \cite{raffel2019exploring}.

\subsection{Abstraction}\label{sub:abstraction}
\begin{figure}[h!]
	\includegraphics[width=0.48\textwidth]{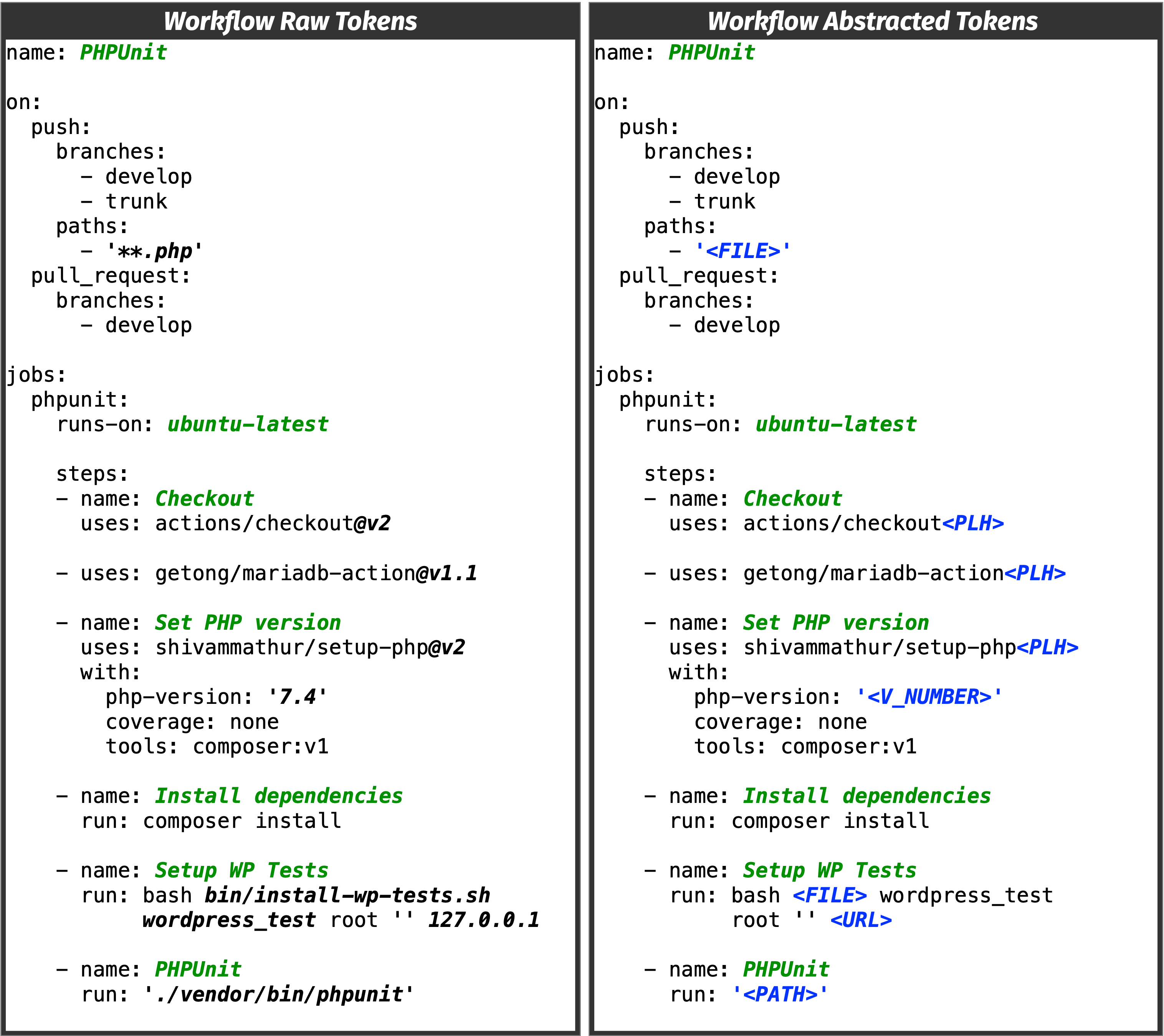}
	\vspace{-0.4cm}
	\caption{Example of Raw and Abstracted Instance.}
	\label{fig:abstracted-raw-example}
	\vspace{-0.5cm}
\end{figure}

We conjecture (and will later experiment) that learning to autocomplete GitHub workflows on raw text (\ie with no preprocessing) is extremely challenging. This is mainly due to the presence of context-specific (and often unique, \ie they have not been seen before) elements in the workflows, such as \texttt{paths} and \texttt{urls}. For example, the left part of \figref{fig:abstracted-raw-example} shows a GitHub workflow featuring elements such as the \texttt{./vendor/bin/phpunit} path or the specific version of an action the user is using (\eg \textit{actions/checkout@v2}), which are likely to hinder the completion learning. 
These are some of the elements we aim at abstracting with special tokens (\eg replacing a path with the \texttt{$<$PATH$>$} tag), as it can be seen in the right part of \figref{fig:abstracted-raw-example}. 

Such an abstraction moves the definition of these context-specific elements from T5 (now only in charge of indicating the need for \eg a \texttt{$<$PATH$>$}) to the developer. We acknowledge that this might imply a slightly higher effort on the developer's side who needs to ``fill the placeholders'' (\ie the special tags) in the prediction. 

To define the abstraction rules, we leverage the unique set of tokens extracted from the workflows of the projects listed in the GitHub actions dataset by Decan \etal \cite{decan2022use}. The dataset features 67,870 GitHub repositories, 29,778 of which use GitHub workflows, and is the one we use to create our training and testing datasets as described in \secref{sub:datasets}.
Given the list in that dataset, we were able to clone 69,040 GitHub repositories, which is more than the 67,870 for which Decan \etal extracted workflow data. From those, we retrieved all GitHub workflows and extracted their ``tokens''. A token can be an action name, a command to run, the option of a command, a path, \etc Out of 10,188,342 unique tokens, 284,463 appear in one workflow, \ie are very specific, confirming our conjecture about the need for abstraction. We randomly selected 1,000 of those tokens for manual inspection. We clustered them based on their ``type'' (\eg \texttt{path}, \texttt{file}). Such a process has been performed by the first author, with the results checked by three other authors. Such a process led to the definition of five categories of context-specific tokens we aim at abstracting: \texttt{url} (\ie a reference to a web resource, such as an IP address), \texttt{file} (\ie a file name/path), \texttt{path} (\ie a path to a directory or to any other resource which cannot be identified as a file since lacking extension), \texttt{version number}, (\ie the specific version of a library, language, or other resources being used), and \texttt{action version} (\ie the specific version of an action that is used). For each category, we defined a special token acting as a placeholder during the abstraction. Note that we distinguish between \texttt{version number} and \texttt{action version} since we assume this could provide additional information to the model which might be useful for the learning.

The abstraction example reported in \figref{fig:abstracted-raw-example} shows how we replace the \texttt{action version} of the token \texttt{actions/\-checkout\-@v2} with the special \texttt{<PLH>} token, while \texttt{files} and \texttt{urls} such as  \texttt{bin/\-install-wp-test.sh} and \texttt{127.0.0.1} are replaced with \texttt{$\langle$FILE$\rangle$} and \texttt{$\langle$URL$\rangle$}, respectively. The code implementing our abstraction process is publicly available \cite{replication}. In a nutshell, we use regular expressions and heuristics to identify the token types of interest and abstract them. 
%
The identification of \texttt{files} leverages, besides a regular expression, a list of extensions we defined during the manual analysis of the tokens appearing in a single workflow. Such a list is also provided in our replication package \cite{replication}.

\rev{To validate our choice of the specific tokens to abstract, we extracted all single-occurring tokens in our dataset, namely those certainly representing problematic cases for any data-driven technique. In total, we identified 23,273 distinct single-occurring tokens. Out of these: 8,226 (37\%) are \texttt{paths}, 8,068 (35\%) are \texttt{files}, 2,833 (12\%) are \texttt{urls}, and 2,334 (10\%) are \texttt{versions}. This means that $\sim$93\% of single-occurring tokens are abstracted by our procedure. This indicates that the proposed abstraction strategy is suitable to abstract rarely-occurring tokens.}

\subsection{Training and Testing Datasets}
\label{sub:datasets}

\subsubsection{Pre-training dataset}
\label{sec:pre-training}
Since the goal of pre-training is to provide T5 with general knowledge about the language(s) of interest, we built a pre-training dataset featuring \texttt{YAML} files (\ie the language used in GitHub workflows), and in particular both general-purpose \texttt{YAML} files as well as those implementing GitHub actions. The former are used for various purposes, \eg CI/CD scripts for other infrastructures (\eg Travis-CI) or other configuration files. 

GitHub actions feature a syntax closer to workflows and therefore would provide further knowledge during pre-training. 

We collected general-purpose \texttt{YAML} files in two steps. First, we searched for \texttt{YAML} files in the 69,040 GitHub repositories we cloned, while excluding those implementing GitHub workflows that we will use to fine-tune the model (\ie those contained in the \texttt{./github/\-workflows/} directory). This resulted in 443,037 general-purpose \texttt{YAML} files. 

To further expand this dataset, we cloned all public non-forked repositories having at least 100 stars and 100 commits, and created in the time range that goes from 2022-25-01 (\ie the day after Decan \etal built their dataset) to 2022-30-09 (the day in which we performed the data collection). The identification of these repositories has been performed using the GitHub search platform by Dabi\'c \etal \cite{ghs}. 

We successfully cloned additional 1,124 GitHub repositories that are not in the dataset by Decan \etal nor are forks of those.
To create the pre-training dataset, which counts a body of 146,006 general-purpose \texttt{YAML} files, we excluded duplicated instances as well as those including non-ASCII tokens and all those having $\#tokens \geq 1024$. Fixing an upper-bound in terms of the number of tokens for the model's input helps in taming the computational cost of training and is a common practice in the literature exploiting DL models to automate code-related tasks \cite{haque2020improved,wang2021context, mastropaolo2021empirical,tufano2021towards,mastropaolo2022automated,ciniselli2021empirical}.


Concerning the \texttt{YAML} files implementing GitHub actions, we collected 13,638 unique examples about the usage of actions from the GitHub Marketplace \cite{market-github}. 
 
The pre-training dataset features 146,066 general-purpose \texttt{YAML} files and 13,638 \texttt{YAML} files implementing GitHub actions. Each instance in the  dataset is a pair featuring (i) a \texttt{YAML} file with 15\% of its tokens randomly masked, and (ii) the expected target, namely the tokens the model is expected to predict instead of the masked ones.

\subsubsection{Fine-tuning dataset}
\label{sec:fine-tuning}
Our fine-tuning dataset features 73,708 GitHub workflows from the whole body of GitHub projects made available by Decan \etal \cite{decan2022use}. On top of those, we mined 733 workflows from the 1,124 GitHub repositories previously mentioned. 

We removed duplicated workflows, and, as done before, all those having $\#tokens \geq 1024$, instances containing non-ASCII characters, and those which overlap with instances in the pre-training dataset. \rev{We were left with 17,935 unique workflows used to train and evaluate \tool. These workflows feature an average of 54 lines (median=41) and 120 tokens (median=84).} 

We split the dataset into training (80\%), validation (10\%), and test (10\%),  making sure that all the instances coming from the same project are assigned to the same subset, thus avoiding leakage of data among the three sets. We obtained 14,348 workflows to train the models, 1,793 for hyperparamenter tuning, and 1,794 to test the best configuration identified. \rev{Each workflow is represented as a JSON-like object preserving the structure of the original workflow file, as it can be seen in the bottom part of \figref{fig:workflow-example}.}

\begin{figure*}[h!]
	\centering
	\includegraphics[width=0.85\linewidth]{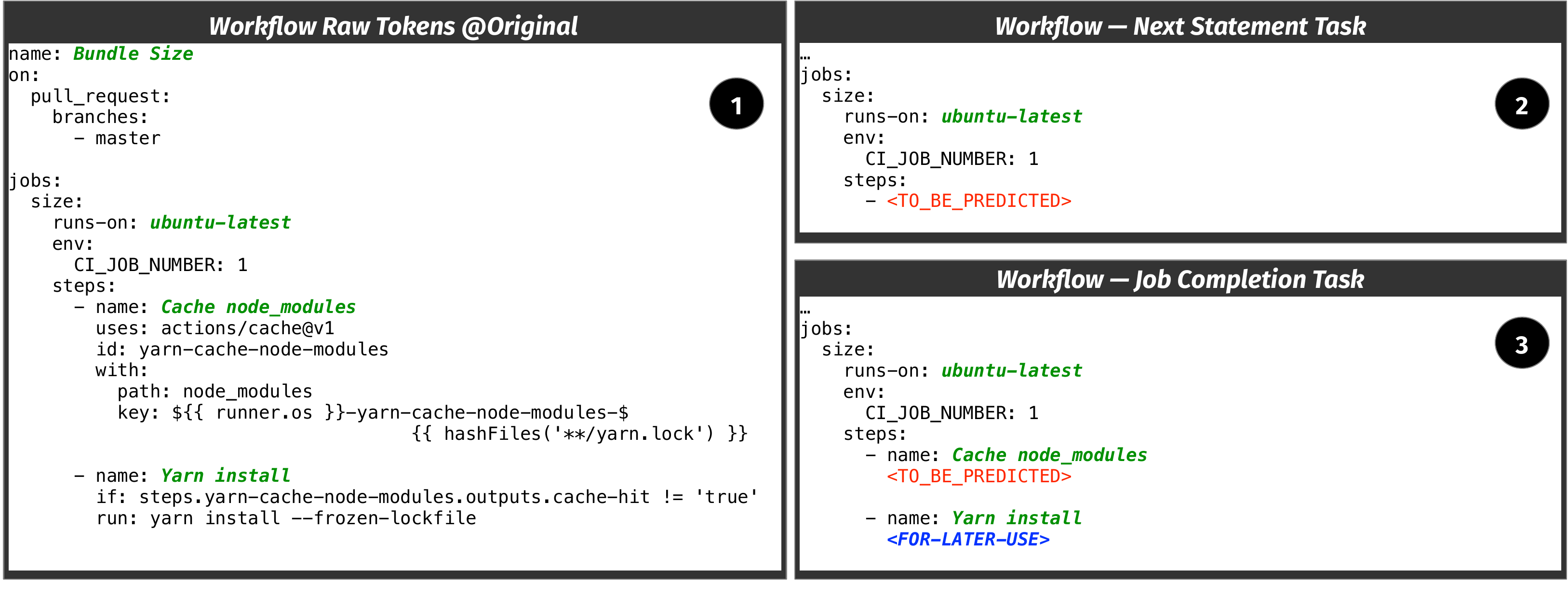}
	\caption{Example of instance for fine-tuning the T5 model on both tasks, namely \cl and \jc}
	\label{fig:workflow-finetuning}
\end{figure*}

We then fine-tune \tool to support two workflow completion scenarios. In the first one, \emph{next step} (\cl), \tool is in charge of predicting the complete $n^{th}$ step a developer is likely to write in a workflow given the preceding already written tokens. A step may or may not contain a textual description (\emph{name}), and it can either consist of action invocations (\emph{uses}) or commands (\emph{run}).
In the second scenario, job completion (\jc), \tool gets as input an abstract job where only \emph{names} are specified, and it is asked to complete it step by step. \figref{fig:workflow-finetuning} helps in better understanding these two scenarios by depicting a fine-tuning instance from our dataset. 

Since we experiment with both the raw workflow version (\ie no abstraction) and with its abstracted version, we report in \figref{fig:workflow-finetuning} an example of ``raw instance''. The left part of the figure \circled{1} shows the original GitHub workflow, while \circled{2} depicts its version for fine-tuning the model for \cl. In this case, we are simulating a scenario in which the developer already wrote the first 11 lines of the workflow (\ie up to \texttt{steps:}), and \tool is asked to predict the first step of the job (\ie \texttt{uses: actions/\-checkout\-@v2}). Note that we can extract multiple (5) training instances from this workflow. Indeed, we can ask the model to predict the first step of the job given just the preceding statements. 

Then, we can ask the model to predict the second step also given the definition of the first step, \etc \figref{fig:workflow-finetuning} \circled{3} depicts a fine-tuning instance for \jc. In this case, we assume that the developer wrote the skeleton of a job by only defining, when available, the job's name it should feature (\eg \emph{Yarn install}). The model is in charge to predict the step masked with the \texttt{<TO\_BE\_PREDICTED>} token, while the \texttt{<FOR-LATER-USE>} token is used to indicate steps that are not yet implemented. Also in this case we can build multiple fine-tuning instances from the workflow in \figref{fig:workflow-finetuning}. We can start predicting the first step in a job using the following $n-1$ for which only the name is provided; then, we can predict the second step, providing the model with the full implementation of the first (as if the model already predicted it) and the following partially defined $n-2$ as context; \etc

\tabref{tab:ds-summary-1} reports the number of instances in the training, validation, and test datasets for both completion scenarios.

\begin{table}[h]
	\centering
	\caption{Number of instances in the used datasets\vspace{-0.3cm}}
	\label{tab:ds-summary-1}
	\begin{tabular}{cccc}
		\toprule
		\textbf{Dataset} & \textbf{train} & \textbf{eval} & \textbf{test}\\ \midrule
		\textit{Pre-training} & 159,645 & -  & - \\
		\textit{Fine-tuning: \cl}  & 108,900     & 13,009  & 13,630 \\
			\textit{Fine-tuning: \jc}               & 108,900   & 13,009 & 13,630     \\
		\bottomrule
	\end{tabular}
	\vspace{-0.3cm}
\end{table}

\subsection{Training and Hyperparameter Tuning} \label{sub:training}
All the trainings we performed have been run using a Google Colab's 2x2, 8 cores TPU topology with a batch size of 32 and an input and target sequence length of 1,350 and 750 tokens, respectively.

\subsubsection{Tokenizer Training}
Since our task is characterized by the presence of natural language and human-readable data-serialization language (\ie \texttt{YAML} data), we trained a new tokenizer (\ie a SentencePiece model \cite{kudo2018sentencepiece} with vocabulary size set to 32k word-pieces) to cope with context-specific elements.
To this extent, we use the 159,645 YAML files included in our pre-training dataset and 712,634 English sentences from the C4 dataset \cite{raffel2019exploring}. The latter is a common practice in literature when developing DL-based models that are required to deal with multi-modal data such as code and technical natural language \cite{mastropaolo2022using,wang2021codet5}.
We included English sentences due to the presence of technical English occurring within GitHub workflows.

\subsubsection{Pre-training strategies}

\rev{We assess \tool in four pre-training scenarios. The first is \emph{No pre-training} (\tfivemodel{NO-PT}), in which the model is not pre-trained, but directly fine-tuned. This means that the model has no previous knowledge of any language and it is just trained to complete GitHub workflows with the available fine-tuning dataset composed by $\sim$109k instances. The second is \emph{YAML pre-training} (\tfivemodel{YL}), in which the model is first pre-trained for 300k steps on a total of 159,645 \texttt{YAML} files including 13,638 actions from the GitHub Marketplace \cite{market-github} and then fine-tuned on the workflow completion task. Thus, in this case the model has knowledge of the general structure of \texttt{YAML} files before being then specialized on the completion task. The third is the \emph{Natural Language Pre-training} (\tfivemodel{NL}), for which we fine-tune the publicly available checkpoint by Raffel \etal \cite{t5-checkpoint} which has been pre-trained for 1M steps on English sentences from the C4 dataset \cite{raffel2019exploring}}. 

\rev{The fourth scenario is \emph{Natural Language+YAML Pre-training} (\tfivemodel{NL+YL}) in which we further pre-trained the previously mentioned checkpoint for additional 300k steps on YAML files, reaching a total of 1,3M pre-training steps (1M on English sentences + 300k on YAML files).}



\subsubsection{Hyperparameter Tuning}
Once pre-trained the models, we fine-tune the hyperparameters of the model following the same procedure employed by Mastropaolo \etal \cite{mastropaolo2021studying}. 

In particular, we assessed the performance of T5 when using four different learning rate schedulers: (i) \textit{Constant Learning Rate} (C-LR): the learning rate is fixed during the whole training; (ii) \textit{Inverse Square Root Learning Rate} (ISR-LR): the learning rate decays as the inverse square root of the training step; (iii) \textit{Slanted Triangular Learning Rate \cite{howard2018universal}} (ST-LR): the learning rate first linearly increases and then linearly decays to the starting learning rate; and (iv) \textit{Polynomial Decay Learning Rate} (PD-LR): the learning rate has a polynomial decay from an initial value to an ending value in the given decay steps. The exact configuration of all the parameters used for each scheduling strategy is reported in our replication package \cite{replication}. Such a procedure has been performed for each of the fine-tuning datasets previously described (\ie both tasks on raw and abstracted code).

%

Having four different training scenarios, four possible learning rates, two different completion contexts, and two versions of the fine-tuning dataset (\ie abstracted and raw tokens), the hyperparameter tuning required building and evaluating 64 models. We fine-tuned each model (\ie each configuration) for 100k steps. Then, we compute the percentage of correct predictions (\ie cases in which the model can correctly generate a recommendation) in the evaluation set. 
\tabref{tab:hp-results} reports the achieved results for each of the 64 models we fine-tuned to find the best-performing configuration (which is reported in boldface).



\begin{table}[h]
	\centering
	\caption{Hyperparameters tuning results}
	\vspace{-0.35cm}
	\scriptsize
	\label{tab:hp-results}
	\begin{tabular}{lrrrrr}
	\toprule
	\multicolumn{6}{c}{{\bf No Pre-training }}\\\midrule
	& \multicolumn{2}{c}{{\bf Raw}} & & \multicolumn{2}{c}{{\bf Abstracted}}\\ \cline{2-3} \cline{5-6}
	& {\bf \cl} & {\bf \jc} & & {\bf \cl} & {\bf \jc}\\\hline
	Constant (C-LR) & 11.06\% & 19.24\% && 13.27\% & 26.73\%\\
	Inverse Square Root (ISQ-LR) & \bf 12.38\% & \bf 21.13\% && \bf 14.21\% & \bf 27.86\%\\
	Slanted Triangular (ST-LR)& 10.13\% & 20.95\% && 12.81\% & 26.65\%\\
	Polynomial Decay (PD-LR)& 10.86\% & 19.01\% && 13.78\% & 25.57\%\\\midrule

	\multicolumn{6}{c}{{\bf YAML Pre-training }}\\\midrule
	& \multicolumn{2}{c}{{\bf Raw}} & & \multicolumn{2}{c}{{\bf Abstracted}}\\ \cline{2-3} \cline{5-6}
	& {\bf \cl} & {\bf \jc} & & {\bf \cl} & {\bf \jc}\\\hline
	Constant (C-LR) & \bf 16.26\% & 25.92\% && 19.05\% & 32.35\%\\
	Inverse Square Root (ISQ-LR)  &15.77\% & 25.47\% && 18.93\% & 31.22\%\\
	Slanted Triangular (ST-LR)& 14.26\% & 24.73\% && 18.05\% & 30.96\%\\
	Polynomial Decay (PD-LR)& 16.15\% & \bf 26.01\% && \bf 19.24\% & \bf 32.81\%\\\midrule
	
	\multicolumn{6}{c}{{\bf English Pre-training \cite{raffel2019exploring}}}\\\midrule
	& \multicolumn{2}{c}{{\bf Raw}} & & \multicolumn{2}{c}{{\bf Abstracted}}\\ \cline{2-3} \cline{5-6}
	& {\bf \cl} & {\bf \jc} & & {\bf \cl} & {\bf \jc}\\\hline
	Constant (C-LR) & 18.35\% & 27.18\% && 22.25\% & 34.02\%\\
	Inverse Square Root (ISQ-LR)  & 18.36\% & 27.10\% && 21.70\% & 33.91\%\\
	Slanted Triangular (ST-LR)& 17.67\% & 26.61\% && 21.70\% & 33.25\%\\
	Polynomial Decay (PD-LR)& \bf 18.46\% & \bf 27.47\% && \bf 22.30\% & \bf 34.12\%\\\midrule
	
	\multicolumn{6}{c}{{\bf YAML+English Pre-training}}\\\midrule
	& \multicolumn{2}{c}{{\bf Raw}} & & \multicolumn{2}{c}{{\bf Abstracted}}\\ \cline{2-3} \cline{5-6}
	& {\bf \cl} & {\bf \jc} & & {\bf \cl} & {\bf \jc}\\\hline
	Constant (C-LR) & 18.06\% & 27.40\% && 21.55\% & 32.91\%\\
	Inverse Square Root (ISQ-LR)  & \bf 18.36\% & \bf 28.17\% && \bf 21.84\% & \bf 34.62\%\\
	Slanted Triangular (ST-LR)& 16.50\% & 25.90\% && 18.88\% & 32.11\%\\
	Polynomial Decay (PD-LR)& 18.28\% & 27.33\% && 21.40\% & 33.36\%\\\midrule
	
\end{tabular} 
\vspace{-0.45cm}
\end{table}

\subsubsection{Fine-tuning}
Once identified the best learning rates to use, we fine-tuned the final models using early stopping to avoid overfitting.
In particular, we save checkpoints every 10k steps using a delta of 0.01, and a patience of 5. This means training the model on the fine-tuning dataset and evaluating its performance on the evaluation set every 10k. The training procedure stops if a gain smaller than the delta (0.01) is observed at each 50k step interval and the best-performing checkpoint up to that training step is selected. Complete data about this process is available in our replication package \cite{replication}.

\subsection{Generating Predictions} \label{sub:predictions}
After the model has been trained, we can generate  predictions for the task we aim at supporting using different decoding schema. To this end, we opted for a greedy decoding strategy \cite{sutskever2014sequence} that generates the recommendation, by selecting at each decoding step the token with the highest probability of appearing in a specific position. Thus, a single prediction is generated for an input sequence.

\section{Study Design} \label{sec:design}

The \emph{goal} of our study is to evaluate \tool. The \emph{quality focus} is \tool's ability to provide correct predictions, as well as predictions that, while differing from the ground truth, could still be valuable for developers. We focus on the two completion scenarios previously described: \cl (mimicking a top-down coding adopted by the developer when writing the workflow statement by statement), and (ii) \jc (helping the developer to complete a job with implementation elements given its textual description). The context consists of the test datasets summarized in \tabref{tab:ds-summary-1}.

The study aims at answering the following research questions:

\newcommand{\rqone}{How difficult it is to automatically complete GitHub workflows as compared to other code completion tasks?}

\newcommand{\rqtwo}{How does \tool perform with different pre-training strategies?}

\newcommand{\rqthree}{How does \tool perform for different prediction scenarios?}

\newcommand{\rqfour}{To what extent ``wrong'' recommendations provided by \tool can be leveraged by developers?}

	
	{\bf RQ$_1$: \emph{\rqtwo}} RQ$_1$ assesses the impact of using different pre-training strategies when completing workflows. We experiment with four pre-training strategies, including the lack of pre-training.
	
	{\bf RQ$_2$: \emph{\rqthree}} RQ$_2$ tests \tool in different prediction scenarios, \ie next statement and job-level contextual completion with and without abstraction. We also implement a statistical language model used as a baseline for comparison.
	
	{\bf RQ$_3$: \emph{\rqfour}}  RQ$_3$ gauges the extent to which ``wrong'' predictions (\ie recommendations different from the expected output) can still be useful to developers and thus worth being integrated into CI/CD pipelines after minor changes.

\subsection{Data Collection and Analysis}




To address RQ$_1$, we run the best-performing configuration for each pre-training strategy and scenario (\cl and \jc) against the test sets (\tabref{tab:ds-summary-1}). Then, we compute the percentage of correct predictions, namely cases in which the models can synthesize completions identical to the expected target (\ie the code written by developers). We further assess the quality of the predictions generated using different pre-training strategies by relying on NLP (Natural Language Processing) metrics such as BLEU \cite{papineni2002bleu} and ROUGE \cite{lin2004rouge}. \smallskip

\textbf{BLEU score} (Bilingual Evaluation Understudy) \cite{papineni2002bleu} measures how similar the candidate (predicted) and reference (oracle) texts are. Given a size $n$, the candidate and reference texts are broken into $n$-grams and the algorithm determines how many $n$-grams of the candidate text appear in the reference text. The BLEU score ranges between 0 (the sequences are completely different) and 1 (the sequences are identical). We use the  BLEU-4 variant as did in previous software engineering papers \cite{wang2021context,watson2020learning,tufano2022using}.\smallskip

\textbf{ROUGE} (Recall-Oriented Understudy for Gisting Evaluation) is a set of metrics for evaluating both automatic summarization of texts and machine translation techniques \cite{lin2004rouge}. ROUGE metrics compare an automatically generated summary or translation with a set of reference summaries (typically, human-produced). We use the ROUGE-L which computes the length of the longest common subsequence between a generated and a reference sentence.\smallskip

To answer RQ$_2$, we first select the best-performing models when supporting the completion of GitHub workflow with and without abstraction in both predictions scenario (\cl and \jc). Later, we assess the quality of the predictions using the same set of metrics (\ie correct predictions, BLEU, and ROUGE score) adopted in RQ$_1$. As there is no previous approach to compare \tool against, we  implemented  a baseline leveraging an $n$-gram model which is a specific actualization of a large class of techniques that assign probabilities to sequences of tokens (\ie Statistical-Language-Model \cite{goldberg2017neural}). To train such a model we use the same set of instances used to fine-tune \tool without, however, any masked part. We experimented with three different values of $n$ (\ie n=3, n=5, and n=7), with $n-1$ being the number of tokens on which the prediction of the next token is based upon. The best value for $n$ ($n=3$) has been found by running the models on the evaluation sets (results in our replication package \cite{replication}). 

The best model has then been run on the same test sets used for \tool's assessment. We do not compare \tool against the $n$-gram when job-level information is provided (\jc), since, by construction, such a technique would not leverage the additional knowledge  provided (\ie it only ``looks'' at the tokens preceding the ones to predict). To explain how predictions are generated with the $3$-gram model, let us assume we are completing a piece of workflow having five tokens $T$, of which the last two are masked (M): $\langle T1,T2,T3,M4,M5 \rangle$. We provide, as input to the model, T2 and T3 to predict M4, obtaining the model prediction P4. Then, we use T3 and P4 to predict M5 obtaining the predicted sentence $\langle T1,T2,T3,P4,P5 \rangle$. While \tool autonomously decides when to stop predicting tokens, this is not the case for the $n$-gram model in our usage scenario. We thus defined two heuristics to stop generating tokens. First, we stop when the $n$-gram model does not generate any output token given the preceding $n$-1. 

Second, we rely on the format in which we represent the instances in our datasets: Each instance is a JSON object and we trained all models to generate as output \texttt{\{target\}}, where the two delimiting curly brackets are the result of our JSON-like representation. Thus, we stop generating tokens when we reach a fully-balanced (\ie valid) JSON object for the test instance to predict (\ie the $n$-gram generated the ``closing'' curly bracket and the latter does not close a curly bracket opened in the predicted code but the JSON-related one). 

We complement the quantitative evaluation by performing statistical tests aimed at assessing whether \tool produces better recommendations as compared to the baseline. We use the McNemar's test \cite{mcnemar} (with is a proportion test for dependent samples)  and Odds Ratios (ORs) on the correct predictions both approaches (\ie \tool and $n$-gram) can generate when evaluated in the \cl completion scenarios, working with both abstracted and raw tokens. We also statistically compare the distribution of the BLEU-4 (computed at statement level) and ROUGE, assuming a significance level of 95\% and using the Wilcoxon signed-rank test \cite{wilcoxon}. The (paired)  Cliff's Delta ($d$) is used as effect size \cite{Gris2005a} and it is considered: negligible for $|d| < 0.10$, small for $0.10 \le |d| < 0.33$, medium for $0.33 \le |d| < 0.474$, and large for $|d| \ge 0.474$ \cite{Gris2005a}. Due to multiple comparisons for both statistical tests, we adjust $p$-values using Holm's correction procedure \cite{Holm1979a}.

As for RQ$_3$, we perform a twofold analysis. We first assess whether the confidence of the model in the generated predictions can be used as a reliable proxy of their ``quality''. T5 provides a \emph{score} for each generated prediction which represents the log-likelihood of the prediction. For example, having a log-likelihood of -2 means that the prediction has a likelihood of 0.69 ($ln(x)=-2 \Longrightarrow x=0.69$). The likelihood can be interpreted as the confidence of the model about the correctness of the prediction on a scale from 0.00 to 1.00 (the higher the better). We split the predictions generated by T5 into ten buckets at steps of 0.1 (\ie the lowest confidence scenario groups the predictions having confidence between 0.0 and 0.1, the highest from 0.9 to 1.0) and report the percentage of correct and wrong predictions within each bucket. Then, given the positive results we achieved (as we will show, the confidence values are representative of the prediction quality), we randomly sample 384 cases of wrong predictions having a confidence $\geq$0.70, with 384 representing a statistically significant sample with a confidence level of 95\% and confidence interval of $\pm5\%$. 

Each sample has been manually classified by two authors with one of the following labels:

\begin{compactenum}
	\item A minor change is required to make the suggestion usable, \eg change an option or a value;
	\item \tool has recommended the correct action/script command, yet with wrong arguments;
	\item \tool has recommended the  correct action/script command, yet with the wrong name;
	\item The suggestion is completely wrong, \ie \tool recommendation is completely different from the ground truth.
\end{compactenum}

In the labeling, the two involved authors achieved a Cohen's kappa \cite{cohen1960coefficient} of 0.72, indicating a \emph{substantial agreement} when measuring inter-rater reliability for categorical items. 

Conflicts, which occurred for 17.97\% of inspected samples, have been solved through open discussion among the authors. 

We report the percentage of predictions assigned to each label and discuss qualitative examples of wrong predictions which, however, might still be valuable for developers. 



\section{Study Results} \label{sec:results}

\textbf{RQ$_1$: \rqtwo} The results obtained by fine-tuning T5 using different pre-training strategies are presented in \tabref{tab:stats-pt-strategies}. The table shows the model's performance in terms of correct predictions, BLEU-4, and ROUGE-LCS (F-measure). The best model for a given combination of task (\ie \cl and \jc) and evaluation metrics is reported in boldface. As expected, the \tfivemodel{NO-PT} is outperformed by all pre-trained models, with 11.23\% and 19.74\% correct predictions for the \cl and \jc task, respectively, when working on raw code. 
When abstracting the dataset, the correct predictions for the \tfivemodel{NO-PT} model improve---14.14\% for \cl and 26.96\% for \jc---while remaining the worst configuration.

\begin{table}[ht]
	\caption{Comparison among different pre-training strategies in terms of correct predictions, BLEU-4 and ROUGE-LCS (f-measure) computed at corpus level}\vspace{-0.3cm}
	\label{tab:stats-pt-strategies}
	 \resizebox{\columnwidth}{!}{%
	\begin{tabular}{llrrrrrr}
		\hline
		\multirow{2}{*}{\bf Dataset} & \multirow{2}{*}{\bf{PT-Strategy}} & \multicolumn{2}{c}{\bf Correct predictions}
		& \multicolumn{2}{c}{\bf BLEU 4} & \multicolumn{2}{c}{\bf ROUGE-LCS} \\
		&&  \textbf{\cl} & \textbf{\jc} 
		& \textbf{\cl} & \textbf{\jc} 
		& \textbf{\cl} & \textbf{\jc}\\
		\midrule
		\multirow{4}{*}{Raw}  
			&  \tfivemodel{NO-PT}& 11.23\% & 19.74\% & 13.70\% & 13.80\% & 44.0\% & 54.75\%\\
			&  \tfivemodel{YL} & 15.85\% & 24.51\% & 14.50\% & 24.10\% & 50.09\% & 61.20\%\\
			& \tfivemodel{NL} \cite{t5-checkpoint}& \bf 17.47\% & 26.02\% & \bf23.10\% & \bf 29.60\% & \bf 51.78\% & 63.34\%\\
			& \tfivemodel{NL+YL} & 17.33\% & \bf 26.35\% & 16.40\% &  27.70\% & 51.74\% & \bf 63.58\%\\
		\hline
		\multirow{4}{*}{Abstracted} 
			&  \tfivemodel{NO-PT} & 14.14\% & 26.98\% & 20.40\% & 24.20\% & 46.31\% & 59.92\%\\ 
			& \tfivemodel{YL} & 19.81\% & 32.58\% & 13.80\% & 17.0\% & 53.30\% & 64.88\% \\ 
			& \tfivemodel{NL}\cite{t5-checkpoint} & 21.28\% & 33.84\%& \bf 28.40\% & \bf 25.90\% & 55.30\% & 66.51\%\\ 
			& \tfivemodel{NL+YL} & \bf 21.36\% & \bf 34.23\% & 21.80\% & 18.40\% & \bf55.37\% & \bf 66.54\%\\ 
	\hline
	\end{tabular}
}
\vspace{-0.2cm}
\end{table}

The results with pre-training (also) involving English documents  (\tfivemodel{NL} and \tfivemodel{NL+YL}) are always the best or the second-best in class, with performance very close to each other. Noteworthy, the usefulness of pre-training on English text when dealing with software-related tasks has been already documented in the literature \cite{tufano2022generating} and is likely due to the vast presence of English terms in the code. Both \tfivemodel{NL} and \tfivemodel{NL+YL} models achieve the best performance on the abstracted workflows, with a percentage of correct predictions of around 21\% for the \cl task and 34\% for the \jc task. 

Two observations can be made here. First, in the \jc task, T5 is more successful thanks to the additional context provided before triggering the prediction (\ie the skeleton of the job defined by the developer---see \secref{sec:fine-tuning}). 

Second, the abstraction seems to substantially boost the model's performance, with $\sim$4\% of additional correct predictions for the \cl task and $\sim$8\% in the \jc task.

\tabref{tab:pretraining-mc} statistically compares the correct predictions achieved using the four different pre-training strategies for the two tasks and the two workflow representations (raw and abstract). Confirming what was said above, the performance of  \tfivemodel{NL} and \tfivemodel{NL+YL}  is always significantly better (adjusted $p$-value $<$ 0.001) compared to the non-pre-trained models (\tfivemodel{NO-PT}) and to the ones pre-trained using \texttt{YAML} files only (\tfivemodel{YL}), with ORs going from 1.49 up to 4.88. The difference between \tfivemodel{NL} and \tfivemodel{NL+YL} is never statistically significant, showing that the two models are almost equivalent. This is an important finding because it means that an English pre-trained model can be simply fine-tuned to successfully accomplish the task (this is way less demanding than retraining the model).

\begin{table}[ht]
\tiny
\caption{Effect of different pre-training strategies on performance: results of McNemar's test.}
\label{tab:pretraining-mc}

\resizebox{\columnwidth}{!}{%
		\begin{tabular}{lllrr}
		  \hline
		\textbf{Dataset} & \textbf{Task} & \textbf{Comparison} & \textbf{\emph{p}-value} & \textbf{OR} \\ 
		  \toprule
		        \multirow{8}{*}{Raw Tokens}   & \multirow{4}{*}{\cl} 
		      	  &\tfivemodel{NL} vs. \tfivemodel{NO-PT} & $<$0.001 & 4.88 \\   
			   & &\tfivemodel{NL} vs. \tfivemodel{YL} & $<$0.001 & 1.95 \\ 
			  &  & \tfivemodel{NL} vs. \tfivemodel{NL+YL} &  0.50 & 1.05 \\ 
			  &  & \tfivemodel{NL+YL} vs \tfivemodel{YL} & $<$0.001 & 1.96 \\ 
		  \cline{2-5}
		  
		  & \multirow{4}{*}{\jc} 
		 	  &\tfivemodel{NL} vs. \tfivemodel{NO-PT} & $<$0.001 & 3.60 \\   
			& &\tfivemodel{NL} vs. \tfivemodel{YL} & $<$0.001 & 1.59 \\ 
			  &  & \tfivemodel{NL} vs. \tfivemodel{NL+YL} &  0.10 & 0.88 \\ 
			&  & \tfivemodel{NL+YL} vs \tfivemodel{YL} & $<$0.001 & 1.74 \\ 
		  \midrule
		   \multirow{8}{*}{Abstracted Tokens} & \multirow{4}{*}{\cl} 
	 	  &\tfivemodel{NL} vs. \tfivemodel{NO-PT} & $<$0.001 & 3.98 \\   
		& &\tfivemodel{NL} vs. \tfivemodel{YL} & $<$0.001 & 1.75 \\ 
		&  & \tfivemodel{NL} vs. \tfivemodel{NL+YL} & 0.69 & 0.96 \\ 
		&  & \tfivemodel{NL+YL} vs \tfivemodel{YL} & $<$0.001 & 1.88 \\ 
		   \cline{2-5}
		   & \multirow{4}{*}{\jc}  
	 	  &\tfivemodel{NL} vs. \tfivemodel{NO-PT} & $<$0.001 & 3.78 \\   
	& &\tfivemodel{NL} vs. \tfivemodel{YL} & $<$0.001 & 1.49 \\ 
	&  & \tfivemodel{NL} vs. \tfivemodel{NL+YL} & 0.05  & 0.86 \\ 
	&  & \tfivemodel{NL+YL} vs \tfivemodel{YL} & $<$0.001 & 1.70 \\ 
		  \bottomrule
		\end{tabular}
}
\vspace{-0.2cm}
\end{table}

The analysis of the BLEU and ROUGE metrics shown in \tabref{tab:stats-pt-strategies} confirms the above-described finding, \ie pre-training always helps, in particular when leveraging English sentences. 
\vspace{0.3cm}
\begin{resultbox}
	\textbf{Answer to RQ$_1$.} The pre-training boosts the performance of \tool. Pre-training with English text (possibly along with YAML files) helps to achieve the best performance. 
\end{resultbox}
\vspace{0.2cm}

In the following RQs we leverage the model pre-trained on English text and \texttt{YAML} files as the backbone of \tool.

\begin{figure*}[h!]
	\centering
	\includegraphics[width=0.94\linewidth]{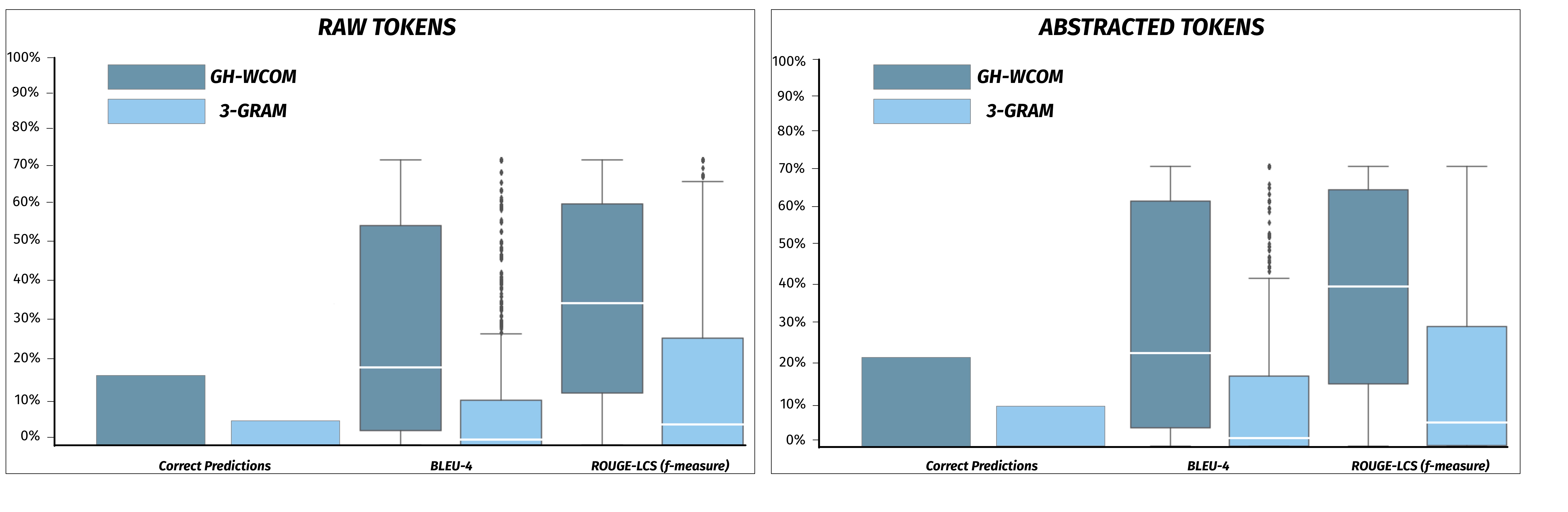}
	\vspace{-0.3cm}
	\caption{Results achieved by \tool and the $n$-gram model when predicting actions for \cl}
	\label{fig:approach}
\end{figure*}

\begin{table}[ht]
	\vspace{-0.1cm}
	\centering
	\caption{\tool vs $3$-gram model when generating recommendations for the \cl }
	\label{tab:stats-approaches}
 \resizebox{\columnwidth}{!}{%
		\begin{tabular}{lllrcc}
			\hline
			\textbf{Dataset}  & \textbf{Comparison} & \textbf{Metric} & \textbf{\emph{p}-value} & \textbf{d} & \textbf{OR}\\ 
			\hline
			
		\multirow{3}{*}{Raw tokens} & \multirow{3}{*}{\tool vs. $n$-gram} & Correct Predictions & $<$0.001 & - & 17.69\\ 
		&& BLEU-4 & $<$0.001 & 0.51 (L) &  - \\ 
		&& ROUGE-LCS & $<$0.001 & 0.52 (L) & - \\\midrule
			
		\multirow{3}{*}{Abstracted tokens} & \multirow{3}{*}{\tool vs. $n$-gram} & Correct Predictions & $<$0.001 & - & 13.76\\ 
	&& BLEU-4 & $<$0.001 & 0.49 (L) &  - \\ 
	&& ROUGE-LCS & $<$0.001 & 0.50 (L) & - \\
			
			\hline
	\end{tabular}
}
\end{table}

\textbf{RQ$_2$: \rqthree} \figref{fig:approach} depicts the results achieved by \tool and the best-performing $n$-gram model ($3$-gram) in terms of correct predictions, BLEU-4 and ROUGE-LCS. Due to the technical limitations of the $n$-gram (\ie it only considers the $n-1$ preceding tokens when generating a prediction), such a comparison has been performed only for the \cl task. 

\tabref{tab:stats-approaches} reports the results of the statistical comparison between the two in terms of adjusted $p$-value and OR (for correct predictions) and effect size (for BLEU and ROUGE). On both datasets, \tool achieves statistically significant better results than the baseline for all metrics. When looking at the correct predictions the gap is of $\sim$11\% on the raw dataset (5.10\% \emph{vs} 17.33\%) and $\sim$12\% on the abstracted dataset (9.28\% \emph{vs} 21.36\%). The OR is 17.69 (raw) and 13.76 (abstract). An OR of 13.76 indicates $\sim$13 times higher odds of obtaining a correct prediction using \tool. Even the comparisons in terms of BLEU and ROUGE show the superiority of \tool both visually (\figref{fig:approach}) and statistically (\tabref{tab:stats-approaches}). 
 
\tool achieves its best performance for the \jc task, with 34.23\% of correct predictions (see \tabref{tab:stats-pt-strategies}), benefiting from the additional contextual information provided as input. Truly, one may question the usefulness of an approach that fails 66\% of the times. Nevertheless, as a term for comparison, the DL-based approach recently proposed by Ciniselli \etal \cite{ciniselli2021empirical} for block-level \java completion achieved $\sim$27\% of correct predictions.
\vspace{0.1cm}
\begin{resultbox}
	\textbf{Answer to RQ$_2$.} \tool outperforms the $n$-gram baseline for the \cl task on all the considered metrics. The gap in correct predictions is $>$11\% on both the raw and the abstracted dataset. The best performances are achieved for the \jc task ($\sim$34\% of correct predictions) thanks to the additional contextual information provided as input.
\end{resultbox}

\begin{figure}[h!]
	
	\includegraphics[width=\columnwidth]{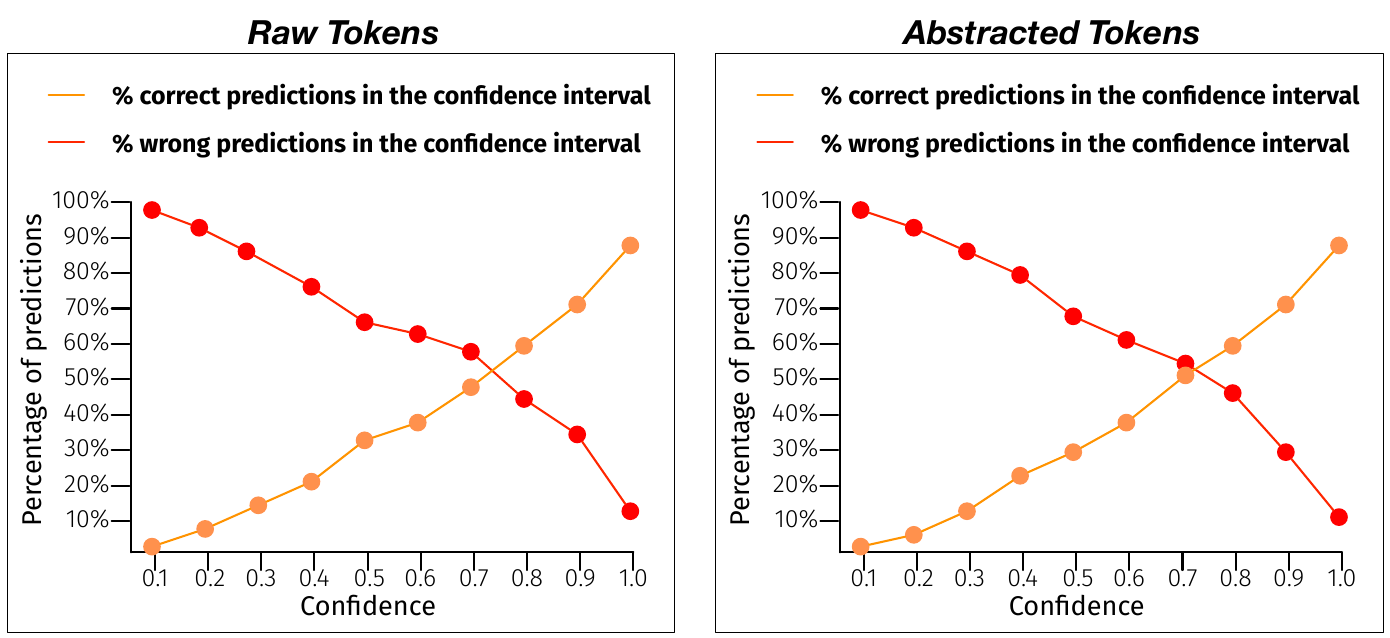}
	\caption{Correct and wrong predictions by the confidence of \tool when generating recommendations for \cl}
	\label{fig:confidence}
\end{figure}

\textbf{RQ$_3$: \rqfour} \figref{fig:confidence} depicts the relationship between the percentage of correct and wrong predictions when considering their confidence. Due to space limitations, we only focus our discussion on the most challenging scenario, namely \cl, as the findings for \jc are similar (complete results in \cite{replication}). The orange line shows the percentage of correct predictions within each confidence interval, \eg 68.45\% of predictions having confidence between 0.8 and 0.9 are correct when working with the raw code. In contrast, the red line shows the percentage of wrong predictions within each confidence bucket. \figref{fig:confidence} shows a clear relationship between the confidence of the predictions and their likelihood of being correct. For example, out of the 1,076 predictions generated with confidence $>$0.9 in the abstracted dataset, 959 (89.13\%) are correct. 

This result has an important practical implication: By setting a threshold on confidence, it would be possible to filter out recommendations likely to be false positives and only notify the developer when the model is quite confident about the generated prediction. As previously said, the results for the \jc are in line with those discussed for \cl. For example, 89.03\% of the 2,908 predictions having confidence $>$0.9 are correct in the abstracted dataset. A similar percentage is achieved on the raw dataset (89.13\%). 

\begin{figure}[h!]
	\includegraphics[width=\columnwidth]{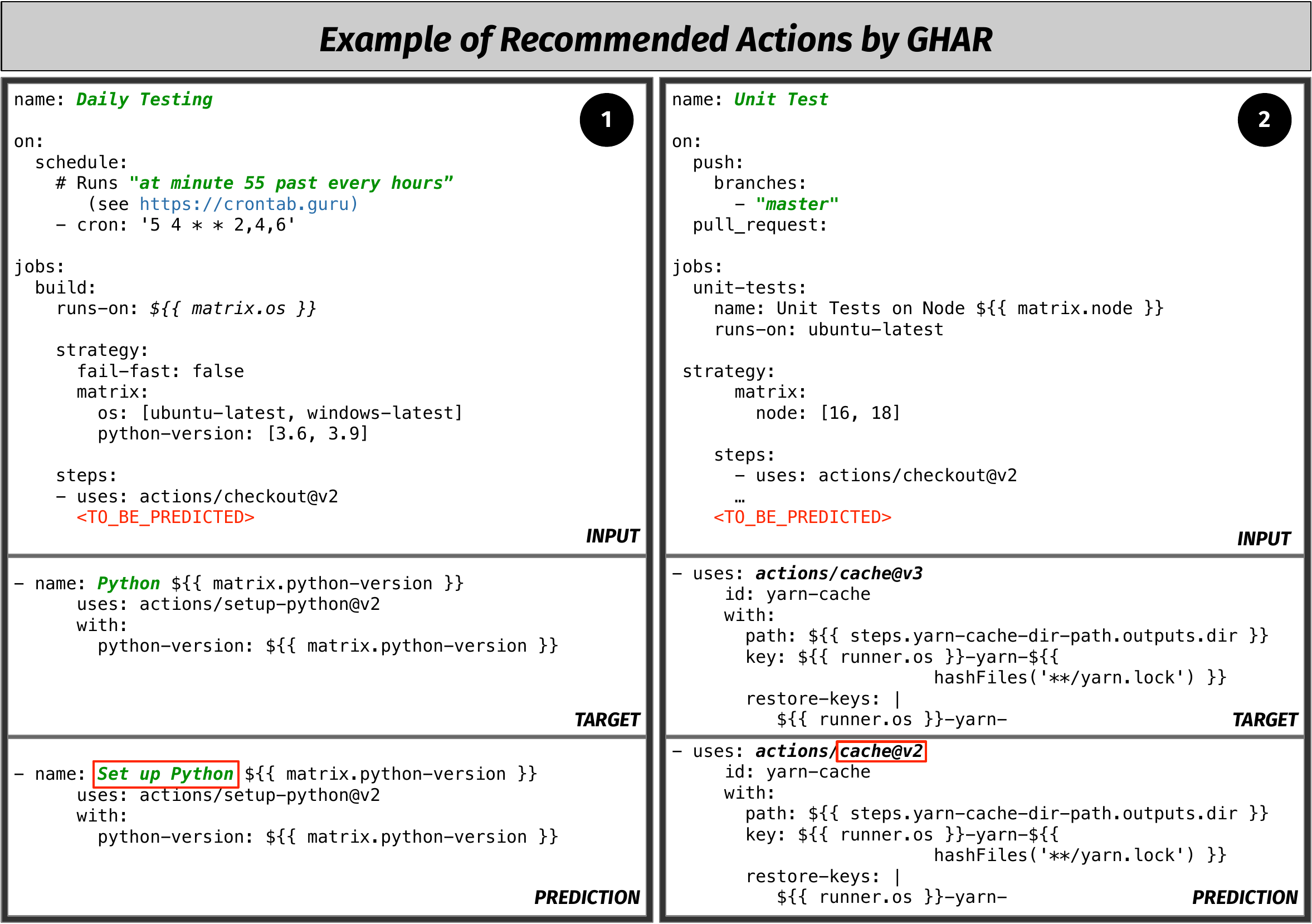}
	\vspace{-0.6cm}
	\caption{Examples of \tool's recommended actions extracted from the manual investigation we performed}
	\label{fig:qualitative-examples}
	\vspace{-0.3cm}
\end{figure}

Concerning the manual analysis of a sample of 384 completions ``wrongly'' predicted by \tool (\ie the prediction did not match the expected target), we found that: (i) 41.41\% (159) are actually wrong, since the predicted code would implement a different behavior than the ground-truth; (ii) in 25.52\% (98) of the cases, \tool suggested the correct action/script command yet with wrong arguments; (iii)  28.13\% (108) of predictions would require minor changes, \rev{implying, on average, changing (\ie insertion and/or deletion)  $\sim$11 characters in the recommended output in order to align with the ground truth;} and (iv) 4.95\% (19) feature a wrong or missing action name, \ie just missing documentation. While the complete results of our manual inspection are available in our replication package \cite{replication}, \figref{fig:qualitative-examples} shows two concrete examples of the instances we inspected. The left part of \figref{fig:qualitative-examples} \circled{1} shows an example in which the whole step is correctly predicted, with the exception of the name which is different from the expected one (\texttt{Set up Python} \emph{vs} \texttt{Python}) but still meaningful. 

The right part \circled{2} depicts a case in which the only difference between the predicted and the expected step is the version of a specific action to use (\texttt{@v2} \emph{vs} \texttt{@v3}). In both cases, the developer is still likely to benefit from the prediction.

\begin{resultbox}
	\textbf{Answer to RQ$_3$.} The confidence of the predictions can serve as a trustworthy indicator of their correctness when auto-completing GitHub workflows; $\sim$50\% of predictions differing from the expected target but on which the model has high confidence could still be valuable for developers.
\end{resultbox}
\vspace{-0.3cm}

\subsection{Why not just using a state-of-the-art chatbot or code recommender?}

Large Language Models (LLMs) have opened up new possibilities even in the field of software engineering. One such application is GitHub Copilot \cite{copilot}, developed by Microsoft using the OpenAI Codex model. Copilot is a state-of-the-art tool for recommending code completion and generation tasks. Similarly, OpenAI's ChatGPT \cite{chatgpt} showed remarkable performance in generating human-like text responses to prompts, even for code-related tasks. 

We conducted a study to investigate the potential of these techniques for supporting auto-completion in GitHub workflows. 
\rev{We tested both tools on 60 instances in our test set by randomly selecting: (i) 15 workflows with the highest confidence score for which GH-WCOM provided correct predictions; (ii) 15 workflows with the highest confidence score for which GH-WCOM failed to provide meaningful recommendations; (iii) 15 workflows with the lowest confidence score for which GH-WCOM provided correct predictions; and (iv) 15 workflows with the lowest confidence score for which GH-WCOM failed to provide meaningful recommendations.}

Concerning the high-confidence scenario, GitHub Copilot was able to provide correct recommendations for 7 of the 15 instances successfully predicted by GH-WCOM. For 2 instances, Copilot did not suggest any token, and for 6 instances, it provided incorrect recommendations. In contrast, when it came to the 15 instances for which \tool generated incorrect recommendations, Copilot correctly recommended only 2 of them and failed to provide meaningful recommendations for the remaining 13.  Regarding ChatGPT, we observed that, out of the 15 instances correctly predicted by \tool, the  chatbot can only suggest 4 meaningful GitHub workflow completions, while providing incorrect action elements/scripts for the remaining 11 instances.  

We then tested ChatGPT on the instances where GH-WCOM failed, we found that for 13 out of 15 workflows, the recommended actions were incorrect, and, for 2 instances, ChatGPT was unable to respond to our query.

\rev{
	As for the GH-WCOM low-confidence instances, also Copilot and ChatGPT poorly performed on such instances. For the 15 successful predictions generated by GH-WCOM, Copilot succeeds in only 4 and ChatGPT in only 3 of them. Copilot and ChatGPT also fail in all 15 cases for which GH-WCOM provides a wrong output. 
}


\vspace{-0.2cm}
\section{Threats to Validity} \label{sec:threats}

\textbf{Construct validity.} \rev{One potential threat arises from the collection of our dataset, as we excluded workflows longer than 1,024 tokens. As mentioned earlier, it is a common practice to limit the input size of DL models to manage training complexity effectively. We recognize that using different thresholds could yield varying results, and we acknowledge this as a potential limitation.}

Another concern involves the extent to which the masking is representative of what programmers do during their tasks \cite{HellendoornPGB19}. We have simulated two scenarios, \cl and \jc, representative of when developers write steps sequentially or code them after sketching their documentation.
To evaluate the quality of the predictions, we used consolidated measures such as the percentage of correct predictions, BLEU-4 \cite{papineni2002bleu,Ren:codebleu}, and ROUGE score. Furthermore, we complemented such measures qualitative analyses. 

\rev{In an attempt to help the model learning, we employed an abstraction schema in which five types of tokens are abstracted with special placeholders. The goal of our abstraction process was to identify a sort of upper-bound for the capabilities of our approach in a \emph{best case} scenario, in which all tokes being \eg a path would be replaced with the same $\langle$PATH$\rangle$ placeholder. Such a simplification pushes more effort on the developer's side while, however, simplifying the learning, and thus representing an upper bound in terms of prediction performances (with the lower bound represented by the raw predictions). We acknowledge that alternative (and less extreme) solutions are possible; for example, distinct paths appearing within the same workflow could be abstracted with different placeholders (\eg $\langle$PATH1$\rangle$, $\langle$PATH2$\rangle$) with the model expected to use the same placeholder for related paths (\ie the same path appearing multiple times in the workflow). As part of our upcoming work agenda, we anticipate conducting user studies to assess different abstraction techniques as alternatives.
}

\textbf{Internal validity.} One key issue for DL models is the hyperparameter tuning, which we detailed in \secref{sec:fine-tuning}. We are aware that we could not consider all possible (combinations of) values for that. Also, the performances of a T5 model could largely depend on how it has been pre-trained. To mitigate this threat, we have shown how \tool works by leveraging different pre-trainings.

\textbf{Conclusion validity.} To address the RQs, wherever appropriate we use suitable statistical tests (McNemar's test and Wilcoxon signed rank test) as well as effect size measures (OR and Cliff's delta). In the qualitative analysis of RQ$_4$, we computed and reported Cohen's kappa inter-rater agreement.

\textbf{External validity.} We experiment \tool with a T5$_{small}$ model. We acknowledge that our choice of the specific model architecture to use could affect the generalizability of our findings. 

For example, larger T5 versions \cite{raffel2019exploring} could lead to different performance. 
\rev{We performed a minimal check of how scaling up the model could affect our findings. To this aim, we trained a T5$_{base}$ model \cite{raffel2019exploring} using the \tfivemodel{NL+YL} setting and the same training process used for T5$_{small}$: We further pre-trained the publicly released T5$_{base}$ checkpoint (pre-trained on natural language) for 300k steps on YAML files and then fine-tuned it on the GitHub workflows. We used the same learning rate scheduler used for T5$_{base}$ (\ie ISQ-LR). The achieved results show that scaling up the model size from 60M to 220M parameters yields negligible improvements in comparison to T5$_{small}$. When employing a T5$_{base}$ architecture to recommend actions in the most demanding scenario (\cl), the difference in correct predictions is a +0.18\% (21.54\%) and a  +0.47\% (17.80\%) for the raw and abstracted datasets, respectively. When incorporating contextual information into the model (\jc), similar conclusions arise (up to +0.67\% of correct predictions).}
  Furthermore, while we applied \tool for GitHub workflow completion, with proper training/fine-tuning, \tool could be applied to CI/CD pipelines developed with different technologies, \eg Jenkins or GitLab. 

\section{Related Work} \label{sec:related}
We discuss literature on automated code completion (which has commonalities with GitHub workflow auto-completion. In particular, we discuss work about task-oriented and pre-trained models.

\subsection{Task-Oriented models for Completing Code}

Li \etal \cite{li2017code} introduce a pointer mixture network improving the accuracy of predicting Out-of-Vocabulary (OoV) words. The pointer mixture network can determine whether to create a word within the vocabulary using an RNN component or reconstruct an OoV word based on the local context using a pointer component. 

Alon \etal \cite{alon2019structural} propose a language-agnostic approach for code completion which uses the syntax to model a code snippet as a tree. Their model predicts the next token in a partial expression represented by an AST, achieving an exact match accuracy of 18.04\%.

Chen \etal \cite{chen2020holistic} focus on recommendaing APIs. Their approach employs a DL technique integrating structural and textual code information with the use of an API context graph and code token network.  Their model outperforms existing graph-based statistical and tree-based DL methods for API recommendation. 

Avishkar \etal \cite{bhoopchand2016learning} propose a neural language model suggesting code in Python using a sparse pointer network to capture long-range relationships among identifiers.
Aye and Kaiser \cite{aye2020sequence} introduce a new language model that predicts the next top-k tokens while taking into account real-world constraints, including prediction latency, model size and memory usage, and suggestion validity. Svyatkovskiy \etal \cite{svyatkovskiy2020fast} propose a learning-to-rank approach for code completion, which is cheaper in terms of memory footprint than generative models. 


\subsection{Pre-trained Models for Code Completion}

Svyatkovskiy \etal~\cite{svyatkovskiy2020intellicode}  introduce IntelliCode, a multilingual code completion tool that predicts sequences of arbitrary token types using subtokens to overcome the OoV problem \cite{tufano2019empirical}. 

Liu \etal~\cite{Liu:ase2020} propose a pre-trained Transformer incorporating two tasks: (i) program understanding and (ii) code generation. The model has been fine-tuned to predict the next code token to write.

Kim \etal \cite{kim2021code} use the Transformer architecture by incorporating the syntactic structure of the code to further advance the state-of-the-art next-token prediction by margins ranging from 14\% to 18\% when compared to previous techniques.

Ciniselli \etal \cite{ciniselli2021empirical} examine the effectiveness of Transformer-based models in completing code with varying degrees of complexity. T5 results to be the best model for recommending code completion across different complexities, with an accuracy of $\sim$29\% when predicting entire code blocks.

Our work shares with the aforementioned ones, and in particular with the one by Ciniselli \etal \cite{ciniselli2021empirical}, the use of transformer architectures, and T5 in particular. That being said, unlike many source code artifacts, a GitHub workflow features several elements that are extremely project-specific, \eg dependencies, configuration files, hardware and software configurations to be tested. As detailed in \secref{sub:abstraction}, this has required a complex abstraction process. Last, but not least, the completion scenarios are different from the ones for the source code. For the former one mainly wants to generate the next statement, block, or code construct. For the latter, elements to generate are either job steps (combinations of natural language descriptions, actions, and script calls) or the implementation of a job specified in terms of its names.

LLMs such as GPT-3 \cite{brown2020language} or GPT-4 \cite{openai2023gpt4} have propelled code completion techniques to new heights. GitHub Copilot \cite{chen2021evaluating} is a prime example of this advancement in the field. On a similar note, OpenAI in November 2022 released ChatGPT \cite{chatgpt}, which showcased remarkable abilities even when dealing with code-related tasks. While we did not use LLMs for feasibility and parsimony reasons, yet, we provide some evidence showing that GitHub workflow completion is a challenging task for them as well. Also, \tool can be evolved to replace T5 with LLMs featuring billions of parameters.


\section{Conclusion and Future Works} \label{sec:conclusion}
This paper tackled the problem of automatically completing CI/CD pipeline scripts, and, in particular, GitHub workflows. We proposed \tool, an approach based on T5 \cite{raffel2019exploring} pre-trained models to automatically recommend workflow completions in different scenarios, \ie predicting the next step (\cl), or filling a workflow job given its textual documentation, \ie the \texttt{names} (\jc).

Our empirical analysis found that (i) leveraging a pre-training involving English text (possibly complemented by \texttt{YAML} files) always helps, (ii) the performance of best models range from 17.47\% (\cl task) and 26.35\% (\jc task) for raw correct predictions, to 21.36\% (\cl) and 34.23\% (\jc)  for abstracted correct predictions; and (iii) the model confidence correlates with the likelihood of generating a correct prediction.
Finally, \tool is competitive for context-sensitive completion tasks when compared to LLM-based tools such as CoPilot \cite{copilot} and ChatGPT \cite{chatgpt}.

Future work aims to experiment with alternative DL models, and, possibly, incorporate developers' feedback in the \tool's learning (\eg by using reinforcement learning).
\section*{Acknowledgments}
Mastropaolo and Bavota are supported by the European Research Council (ERC) under the European Union's Horizon 2020 research and innovation programme (grant agreement No. 851720). Mastropaolo thanks CHOOSE for sponsoring his trip to the conference. Zampetti and Di Penta are supported by the Horizon 2020 (EU Commission) project COSMOS (DevOps for Complex Cyber-physical Systems), Project No. 957254-COSMOS.

\bibliographystyle{ACM-Reference-Format}
\bibliography{main}


\begin{thebibliography}{62}


\ifx \showCODEN    \undefined \def \showCODEN     #1{\unskip}     \fi
\ifx \showDOI      \undefined \def \showDOI       #1{#1}\fi
\ifx \showISBNx    \undefined \def \showISBNx     #1{\unskip}     \fi
\ifx \showISBNxiii \undefined \def \showISBNxiii  #1{\unskip}     \fi
\ifx \showISSN     \undefined \def \showISSN      #1{\unskip}     \fi
\ifx \showLCCN     \undefined \def \showLCCN      #1{\unskip}     \fi
\ifx \shownote     \undefined \def \shownote      #1{#1}          \fi
\ifx \showarticletitle \undefined \def \showarticletitle #1{#1}   \fi
\ifx \showURL      \undefined \def \showURL       {\relax}        \fi
\providecommand\bibfield[2]{#2}
\providecommand\bibinfo[2]{#2}
\providecommand\natexlab[1]{#1}
\providecommand\showeprint[2][]{arXiv:#2}

\bibitem[cha({[n.\,d.]})]%
        {chatgpt}
 \bibinfo{year}{[n.\,d.]}\natexlab{}.
\newblock \bibinfo{title}{ChatGPT~\url{https://openai.com/blog/chatgpt}}.
\newblock
\newblock


\bibitem[cop({[n.\,d.]})]%
        {copilot}
 \bibinfo{year}{[n.\,d.]}\natexlab{}.
\newblock \bibinfo{title}{GitHub Copilot~\url{https://copilot.github.com}}.
\newblock
\newblock


\bibitem[mar({[n.\,d.]})]%
        {market-github}
 \bibinfo{year}{[n.\,d.]}\natexlab{}.
\newblock \bibinfo{title}{GitHub Marketplace
  ~\url{https://github.com/marketplace?type=actions}}.
\newblock
\newblock


\bibitem[git({[n.\,d.]})]%
        {github-workflows}
 \bibinfo{year}{[n.\,d.]}\natexlab{}.
\newblock \bibinfo{title}{{GitHub} workflows}.
\newblock
\newblock
\urldef\tempurl%
\url{https://docs.github.com/en/actions/using-workflows}
\showURL{%
\tempurl}
\newblock
\shownote{Last accessed Feb 16, 2023}.


\bibitem[ghs({[n.\,d.]})]%
        {ghs}
 \bibinfo{year}{[n.\,d.]}\natexlab{}.
\newblock \bibinfo{title}{MSR mining platform}.
\newblock \bibinfo{howpublished}{\url{https://seart-ghs.si.usi.ch}}.
\newblock


\bibitem[rep({[n.\,d.]})]%
        {replication}
 \bibinfo{year}{[n.\,d.]}\natexlab{}.
\newblock \bibinfo{title}{Replication
  Package~\url{https://github.com/antonio-mastropaolo/GH-WCOM}}.
\newblock
\newblock


\bibitem[t5-({[n.\,d.]})]%
        {t5-checkpoint}
 \bibinfo{year}{[n.\,d.]}\natexlab{}.
\newblock \bibinfo{title}{T5 public
  checkpoint~\url{gs://t5-data/pretrained_models/small}}.
\newblock
\newblock


\bibitem[Alon et~al\mbox{.}(2020)]%
        {alon2019structural}
\bibfield{author}{\bibinfo{person}{Uri Alon}, \bibinfo{person}{Roy Sadaka},
  \bibinfo{person}{Omer Levy}, {and} \bibinfo{person}{Eran Yahav}.}
  \bibinfo{year}{2020}\natexlab{}.
\newblock \showarticletitle{Structural language models of code}. In
  \bibinfo{booktitle}{\emph{International Conference on Machine Learning}}.
  PMLR, \bibinfo{pages}{245--256}.
\newblock


\bibitem[Aye and Kaiser(2020)]%
        {aye2020sequence}
\bibfield{author}{\bibinfo{person}{Gareth~Ari Aye} {and}
  \bibinfo{person}{Gail~E Kaiser}.} \bibinfo{year}{2020}\natexlab{}.
\newblock \showarticletitle{Sequence Model Design for Code Completion in the
  Modern IDE}.
\newblock \bibinfo{journal}{\emph{arXiv preprint arXiv:2004.05249}}
  (\bibinfo{year}{2020}).
\newblock


\bibitem[Bhoopchand et~al\mbox{.}(2016)]%
        {bhoopchand2016learning}
\bibfield{author}{\bibinfo{person}{Avishkar Bhoopchand}, \bibinfo{person}{Tim
  Rockt{\"a}schel}, \bibinfo{person}{Earl Barr}, {and}
  \bibinfo{person}{Sebastian Riedel}.} \bibinfo{year}{2016}\natexlab{}.
\newblock \showarticletitle{Learning python code suggestion with a sparse
  pointer network}.
\newblock \bibinfo{journal}{\emph{arXiv preprint arXiv:1611.08307}}
  (\bibinfo{year}{2016}).
\newblock


\bibitem[Brown et~al\mbox{.}(2020)]%
        {brown2020language}
\bibfield{author}{\bibinfo{person}{Tom Brown}, \bibinfo{person}{Benjamin Mann},
  \bibinfo{person}{Nick Ryder}, \bibinfo{person}{Melanie Subbiah},
  \bibinfo{person}{Jared~D Kaplan}, \bibinfo{person}{Prafulla Dhariwal},
  \bibinfo{person}{Arvind Neelakantan}, \bibinfo{person}{Pranav Shyam},
  \bibinfo{person}{Girish Sastry}, \bibinfo{person}{Amanda Askell},
  {et~al\mbox{.}}} \bibinfo{year}{2020}\natexlab{}.
\newblock \showarticletitle{Language models are few-shot learners}.
\newblock \bibinfo{journal}{\emph{Advances in neural information processing
  systems}}  \bibinfo{volume}{33} (\bibinfo{year}{2020}),
  \bibinfo{pages}{1877--1901}.
\newblock


\bibitem[Chen et~al\mbox{.}(2021b)]%
        {chen2020holistic}
\bibfield{author}{\bibinfo{person}{Chi Chen}, \bibinfo{person}{Xin Peng},
  \bibinfo{person}{Zhenchang Xing}, \bibinfo{person}{Jun Sun},
  \bibinfo{person}{Xin Wang}, \bibinfo{person}{Yifan Zhao}, {and}
  \bibinfo{person}{Wenyun Zhao}.} \bibinfo{year}{2021}\natexlab{b}.
\newblock \showarticletitle{Holistic combination of structural and textual code
  information for context based API recommendation}.
\newblock \bibinfo{journal}{\emph{IEEE Transactions on Software Engineering}}
  (\bibinfo{year}{2021}).
\newblock


\bibitem[Chen(2015)]%
        {Chen:2015}
\bibfield{author}{\bibinfo{person}{L. Chen}.} \bibinfo{year}{2015}\natexlab{}.
\newblock \showarticletitle{Continuous Delivery: Huge Benefits, but Challenges
  Too}.
\newblock \bibinfo{journal}{\emph{IEEE Software}} \bibinfo{volume}{32},
  \bibinfo{number}{2} (\bibinfo{year}{2015}), \bibinfo{pages}{50--54}.
\newblock


\bibitem[Chen(2017)]%
        {CHEN201772}
\bibfield{author}{\bibinfo{person}{Lianping Chen}.}
  \bibinfo{year}{2017}\natexlab{}.
\newblock \showarticletitle{Continuous Delivery: Overcoming adoption
  challenges}.
\newblock \bibinfo{journal}{\emph{Journal of Systems and Software}}
  \bibinfo{volume}{128} (\bibinfo{year}{2017}), \bibinfo{pages}{72 -- 86}.
\newblock


\bibitem[Chen et~al\mbox{.}(2021c)]%
        {chen2021evaluating}
\bibfield{author}{\bibinfo{person}{Mark Chen}, \bibinfo{person}{Jerry Tworek},
  \bibinfo{person}{Heewoo Jun}, \bibinfo{person}{Qiming Yuan},
  \bibinfo{person}{Henrique Ponde de~Oliveira Pinto}, \bibinfo{person}{Jared
  Kaplan}, \bibinfo{person}{Harri Edwards}, \bibinfo{person}{Yuri Burda},
  \bibinfo{person}{Nicholas Joseph}, \bibinfo{person}{Greg Brockman},
  {et~al\mbox{.}}} \bibinfo{year}{2021}\natexlab{c}.
\newblock \showarticletitle{Evaluating large language models trained on code}.
\newblock \bibinfo{journal}{\emph{arXiv preprint arXiv:2107.03374}}
  (\bibinfo{year}{2021}).
\newblock


\bibitem[Chen et~al\mbox{.}(2023)]%
        {ChenKM23}
\bibfield{author}{\bibinfo{person}{Zimin Chen}, \bibinfo{person}{Steve
  Kommrusch}, {and} \bibinfo{person}{Martin Monperrus}.}
  \bibinfo{year}{2023}\natexlab{}.
\newblock \showarticletitle{Neural Transfer Learning for Repairing Security
  Vulnerabilities in {C} Code}.
\newblock \bibinfo{journal}{\emph{{IEEE} Trans. Software Eng.}}
  \bibinfo{volume}{49}, \bibinfo{number}{1} (\bibinfo{year}{2023}),
  \bibinfo{pages}{147--165}.
\newblock


\bibitem[Chen et~al\mbox{.}(2021a)]%
        {ChenKTPPM21}
\bibfield{author}{\bibinfo{person}{Zimin Chen}, \bibinfo{person}{Steve
  Kommrusch}, \bibinfo{person}{Michele Tufano},
  \bibinfo{person}{Louis{-}No{\"{e}}l Pouchet}, \bibinfo{person}{Denys
  Poshyvanyk}, {and} \bibinfo{person}{Martin Monperrus}.}
  \bibinfo{year}{2021}\natexlab{a}.
\newblock \showarticletitle{SequenceR: Sequence-to-Sequence Learning for
  End-to-End Program Repair}.
\newblock \bibinfo{journal}{\emph{{IEEE} Trans. Software Eng.}}
  \bibinfo{volume}{47}, \bibinfo{number}{9} (\bibinfo{year}{2021}),
  \bibinfo{pages}{1943--1959}.
\newblock


\bibitem[Ciniselli et~al\mbox{.}(2022)]%
        {ciniselli2021empirical}
\bibfield{author}{\bibinfo{person}{Matteo Ciniselli}, \bibinfo{person}{Nathan
  Cooper}, \bibinfo{person}{Luca Pascarella}, \bibinfo{person}{Antonio
  Mastropaolo}, \bibinfo{person}{Emad Aghajani}, \bibinfo{person}{Denys
  Poshyvanyk}, \bibinfo{person}{Massimiliano {Di Penta}}, {and}
  \bibinfo{person}{Gabriele Bavota}.} \bibinfo{year}{2022}\natexlab{}.
\newblock \showarticletitle{An Empirical Study on the Usage of Transformer
  Models for Code Completion}.
\newblock \bibinfo{journal}{\emph{{IEEE} Trans. Software Eng.}}
  \bibinfo{volume}{48}, \bibinfo{number}{12} (\bibinfo{year}{2022}),
  \bibinfo{pages}{4818--4837}.
\newblock
\urldef\tempurl%
\url{https://doi.org/10.1109/TSE.2021.3128234}
\showDOI{\tempurl}


\bibitem[Cohen(1960)]%
        {cohen1960coefficient}
\bibfield{author}{\bibinfo{person}{Jacob Cohen}.}
  \bibinfo{year}{1960}\natexlab{}.
\newblock \showarticletitle{A coefficient of agreement for nominal scales}.
\newblock \bibinfo{journal}{\emph{Educational and psychological measurement}}
  \bibinfo{volume}{20}, \bibinfo{number}{1} (\bibinfo{year}{1960}),
  \bibinfo{pages}{37--46}.
\newblock


\bibitem[Decan et~al\mbox{.}(2022a)]%
        {DecanMMG22}
\bibfield{author}{\bibinfo{person}{Alexandre Decan}, \bibinfo{person}{Tom
  Mens}, \bibinfo{person}{Pooya~Rostami Mazrae}, {and} \bibinfo{person}{Mehdi
  Golzadeh}.} \bibinfo{year}{2022}\natexlab{a}.
\newblock \showarticletitle{On the Use of GitHub Actions in Software
  Development Repositories}. In \bibinfo{booktitle}{\emph{{IEEE} International
  Conference on Software Maintenance and Evolution, {ICSME} 2022, Limassol,
  Cyprus, October 3-7, 2022}}. \bibinfo{publisher}{{IEEE}},
  \bibinfo{pages}{235--245}.
\newblock


\bibitem[Decan et~al\mbox{.}(2022b)]%
        {decan2022use}
\bibfield{author}{\bibinfo{person}{Alexandre Decan}, \bibinfo{person}{Tom
  Mens}, \bibinfo{person}{Pooya~Rostami Mazrae}, {and} \bibinfo{person}{Mehdi
  Golzadeh}.} \bibinfo{year}{2022}\natexlab{b}.
\newblock \showarticletitle{On the Use of GitHub Actions in Software
  Development Repositories}. In \bibinfo{booktitle}{\emph{2022 IEEE
  International Conference on Software Maintenance and Evolution (ICSME)}}.
  IEEE, \bibinfo{pages}{235--245}.
\newblock


\bibitem[Fu et~al\mbox{.}(2022)]%
        {FuTLNP22}
\bibfield{author}{\bibinfo{person}{Michael Fu}, \bibinfo{person}{Chakkrit
  Tantithamthavorn}, \bibinfo{person}{Trung Le}, \bibinfo{person}{Van Nguyen},
  {and} \bibinfo{person}{Dinh~Q. Phung}.} \bibinfo{year}{2022}\natexlab{}.
\newblock \showarticletitle{VulRepair: a {T5}-based automated software
  vulnerability repair}. In \bibinfo{booktitle}{\emph{Proceedings of the 30th
  {ACM} Joint European Software Engineering Conference and Symposium on the
  Foundations of Software Engineering, {ESEC/FSE} 2022, Singapore, Singapore,
  November 14-18, 2022}}. \bibinfo{publisher}{{ACM}},
  \bibinfo{pages}{935--947}.
\newblock


\bibitem[Goldberg(2017)]%
        {goldberg2017neural}
\bibfield{author}{\bibinfo{person}{Yoav Goldberg}.}
  \bibinfo{year}{2017}\natexlab{}.
\newblock \bibinfo{booktitle}{\emph{Neural network methods in natural language
  processing}}.
\newblock \bibinfo{publisher}{Morgan \& Claypool Publishers}.
\newblock


\bibitem[Grissom and Kim(2005)]%
        {Gris2005a}
\bibfield{author}{\bibinfo{person}{Robert~J Grissom} {and}
  \bibinfo{person}{John~J Kim}.} \bibinfo{year}{2005}\natexlab{}.
\newblock \bibinfo{booktitle}{\emph{Effect sizes for research: A broad
  practical approach.}}
\newblock \bibinfo{publisher}{Lawrence Erlbaum Associates Publishers}.
\newblock


\bibitem[Haque et~al\mbox{.}(2020)]%
        {haque2020improved}
\bibfield{author}{\bibinfo{person}{Sakib Haque}, \bibinfo{person}{Alexander
  LeClair}, \bibinfo{person}{Lingfei Wu}, {and} \bibinfo{person}{Collin
  McMillan}.} \bibinfo{year}{2020}\natexlab{}.
\newblock \showarticletitle{Improved automatic summarization of subroutines via
  attention to file context}. In \bibinfo{booktitle}{\emph{Proceedings of the
  17th International Conference on Mining Software Repositories}}.
  \bibinfo{pages}{300--310}.
\newblock


\bibitem[Hellendoorn et~al\mbox{.}(2019)]%
        {HellendoornPGB19}
\bibfield{author}{\bibinfo{person}{Vincent~J. Hellendoorn},
  \bibinfo{person}{Sebastian Proksch}, \bibinfo{person}{Harald~C. Gall}, {and}
  \bibinfo{person}{Alberto Bacchelli}.} \bibinfo{year}{2019}\natexlab{}.
\newblock \showarticletitle{When code completion fails: a case study on
  real-world completions}. In \bibinfo{booktitle}{\emph{Proceedings of the 41st
  International Conference on Software Engineering, {ICSE} 2019, Montreal, QC,
  Canada, May 25-31, 2019}}. \bibinfo{pages}{960--970}.
\newblock


\bibitem[Hilton et~al\mbox{.}(2017)]%
        {HiltonFSE17}
\bibfield{author}{\bibinfo{person}{Michael Hilton}, \bibinfo{person}{Nicholas
  Nelson}, \bibinfo{person}{Timothy Tunnell}, \bibinfo{person}{Darko Marinov},
  {and} \bibinfo{person}{Danny Dig}.} \bibinfo{year}{2017}\natexlab{}.
\newblock \showarticletitle{Trade-Offs in Continuous Integration: Assurance,
  Security, and Flexibility}. In \bibinfo{booktitle}{\emph{Proceedings of the
  25th {ACM} {SIGSOFT} International Symposium on Foundations of Software
  Engineering, {FSE} 2017}}.
\newblock


\bibitem[Holm(1979)]%
        {Holm1979a}
\bibfield{author}{\bibinfo{person}{Sture Holm}.}
  \bibinfo{year}{1979}\natexlab{}.
\newblock \showarticletitle{A simple sequentially rejective multiple test
  procedure}.
\newblock \bibinfo{journal}{\emph{Scandinavian journal of statistics}}
  (\bibinfo{year}{1979}), \bibinfo{pages}{65--70}.
\newblock


\bibitem[Howard and Ruder(2018)]%
        {howard2018universal}
\bibfield{author}{\bibinfo{person}{Jeremy Howard} {and}
  \bibinfo{person}{Sebastian Ruder}.} \bibinfo{year}{2018}\natexlab{}.
\newblock \showarticletitle{Universal language model fine-tuning for text
  classification}.
\newblock \bibinfo{journal}{\emph{arXiv preprint arXiv:1801.06146}}
  (\bibinfo{year}{2018}).
\newblock


\bibitem[Kim et~al\mbox{.}(2021)]%
        {kim2021code}
\bibfield{author}{\bibinfo{person}{Seohyun Kim}, \bibinfo{person}{Jinman Zhao},
  \bibinfo{person}{Yuchi Tian}, {and} \bibinfo{person}{Satish Chandra}.}
  \bibinfo{year}{2021}\natexlab{}.
\newblock \showarticletitle{Code prediction by feeding trees to transformers}.
  In \bibinfo{booktitle}{\emph{2021 IEEE/ACM 43rd International Conference on
  Software Engineering (ICSE)}}. IEEE, \bibinfo{pages}{150--162}.
\newblock


\bibitem[Kudo and Richardson(2018)]%
        {kudo2018sentencepiece}
\bibfield{author}{\bibinfo{person}{Taku Kudo} {and} \bibinfo{person}{John
  Richardson}.} \bibinfo{year}{2018}\natexlab{}.
\newblock \showarticletitle{Sentencepiece: A simple and language independent
  subword tokenizer and detokenizer for neural text processing}.
\newblock \bibinfo{journal}{\emph{arXiv preprint arXiv:1808.06226}}
  (\bibinfo{year}{2018}).
\newblock


\bibitem[Li et~al\mbox{.}(2017)]%
        {li2017code}
\bibfield{author}{\bibinfo{person}{Jian Li}, \bibinfo{person}{Yue Wang},
  \bibinfo{person}{Michael~R Lyu}, {and} \bibinfo{person}{Irwin King}.}
  \bibinfo{year}{2017}\natexlab{}.
\newblock \showarticletitle{Code completion with neural attention and pointer
  networks}.
\newblock \bibinfo{journal}{\emph{arXiv preprint arXiv:1711.09573}}
  (\bibinfo{year}{2017}).
\newblock


\bibitem[Li et~al\mbox{.}(2020)]%
        {li2020repair}
\bibfield{author}{\bibinfo{person}{Yi Li}, \bibinfo{person}{Shaohua Wang},
  {and} \bibinfo{person}{Tien~N. Nguyen}.} \bibinfo{year}{2020}\natexlab{}.
\newblock \showarticletitle{DLFix: Context-Based Code Transformation Learning
  for Automated Program Repair}. In \bibinfo{booktitle}{\emph{Proceedings of
  the ACM/IEEE 42nd International Conference on Software Engineering}} (Seoul,
  South Korea) \emph{(\bibinfo{series}{ICSE '20})}.
  \bibinfo{publisher}{Association for Computing Machinery},
  \bibinfo{address}{New York, NY, USA}, \bibinfo{pages}{602–614}.
\newblock
\showISBNx{9781450371216}
\urldef\tempurl%
\url{https://doi.org/10.1145/3377811.3380345}
\showDOI{\tempurl}


\bibitem[Li et~al\mbox{.}(2022)]%
        {li2022repair}
\bibfield{author}{\bibinfo{person}{Yi Li}, \bibinfo{person}{Shaohua Wang},
  {and} \bibinfo{person}{Tien~N. Nguyen}.} \bibinfo{year}{2022}\natexlab{}.
\newblock \showarticletitle{DEAR: A Novel Deep Learning-Based Approach for
  Automated Program Repair}. In \bibinfo{booktitle}{\emph{Proceedings of the
  44th International Conference on Software Engineering}} (Pittsburgh,
  Pennsylvania) \emph{(\bibinfo{series}{ICSE '22})}.
  \bibinfo{publisher}{Association for Computing Machinery},
  \bibinfo{address}{New York, NY, USA}, \bibinfo{pages}{511–523}.
\newblock
\showISBNx{9781450392211}
\urldef\tempurl%
\url{https://doi.org/10.1145/3510003.3510177}
\showDOI{\tempurl}


\bibitem[Lin(2004)]%
        {lin2004rouge}
\bibfield{author}{\bibinfo{person}{Chin-Yew Lin}.}
  \bibinfo{year}{2004}\natexlab{}.
\newblock \showarticletitle{Rouge: A package for automatic evaluation of
  summaries}. In \bibinfo{booktitle}{\emph{Text summarization branches out}}.
  \bibinfo{pages}{74--81}.
\newblock


\bibitem[Liu et~al\mbox{.}(2020)]%
        {Liu:ase2020}
\bibfield{author}{\bibinfo{person}{Fang Liu}, \bibinfo{person}{Ge Li},
  \bibinfo{person}{Yunfei Zhao}, {and} \bibinfo{person}{Zhi Jin}.}
  \bibinfo{year}{2020}\natexlab{}.
\newblock \showarticletitle{Multi-task Learning based Pre-trained Language
  Model for Code Completion}. In \bibinfo{booktitle}{\emph{Proceedings of the
  35th IEEE/ACM International Conference on Automated Software Engineering}}
  \emph{(\bibinfo{series}{ASE 2020})}. \bibinfo{publisher}{Association for
  Computing Machinery}.
\newblock


\bibitem[Mastropaolo et~al\mbox{.}(2021a)]%
        {mastropaolo2021empirical}
\bibfield{author}{\bibinfo{person}{Antonio Mastropaolo}, \bibinfo{person}{Emad
  Aghajani}, \bibinfo{person}{Luca Pascarella}, {and} \bibinfo{person}{Gabriele
  Bavota}.} \bibinfo{year}{2021}\natexlab{a}.
\newblock \showarticletitle{An Empirical Study on Code Comment Completion}. In
  \bibinfo{booktitle}{\emph{2021 IEEE International Conference on Software
  Maintenance and Evolution (ICSME)}}. IEEE, \bibinfo{pages}{159--170}.
\newblock


\bibitem[Mastropaolo et~al\mbox{.}(2022a)]%
        {mastropaolo2022automated}
\bibfield{author}{\bibinfo{person}{Antonio Mastropaolo}, \bibinfo{person}{Emad
  Aghajani}, \bibinfo{person}{Luca Pascarella}, {and} \bibinfo{person}{Gabriele
  Bavota}.} \bibinfo{year}{2022}\natexlab{a}.
\newblock \showarticletitle{Automated Variable Renaming: Are We There Yet?}
\newblock \bibinfo{journal}{\emph{arXiv preprint arXiv:2212.05738}}
  (\bibinfo{year}{2022}).
\newblock


\bibitem[Mastropaolo et~al\mbox{.}(2022b)]%
        {mastropaolo2022using}
\bibfield{author}{\bibinfo{person}{Antonio Mastropaolo}, \bibinfo{person}{Luca
  Pascarella}, {and} \bibinfo{person}{Gabriele Bavota}.}
  \bibinfo{year}{2022}\natexlab{b}.
\newblock \showarticletitle{Using Deep Learning to Generate Complete Log
  Statements}.
\newblock \bibinfo{journal}{\emph{arXiv preprint arXiv:2201.04837}}
  (\bibinfo{year}{2022}).
\newblock


\bibitem[Mastropaolo et~al\mbox{.}(2021b)]%
        {mastropaolo2021studying}
\bibfield{author}{\bibinfo{person}{Antonio Mastropaolo},
  \bibinfo{person}{Simone Scalabrino}, \bibinfo{person}{Nathan Cooper},
  \bibinfo{person}{David~Nader Palacio}, \bibinfo{person}{Denys Poshyvanyk},
  \bibinfo{person}{Rocco Oliveto}, {and} \bibinfo{person}{Gabriele Bavota}.}
  \bibinfo{year}{2021}\natexlab{b}.
\newblock \showarticletitle{Studying the usage of text-to-text transfer
  transformer to support code-related tasks}. In \bibinfo{booktitle}{\emph{2021
  IEEE/ACM 43rd International Conference on Software Engineering (ICSE)}}.
  IEEE, \bibinfo{pages}{336--347}.
\newblock


\bibitem[McNemar(1947)]%
        {mcnemar}
\bibfield{author}{\bibinfo{person}{Quinn McNemar}.}
  \bibinfo{year}{1947}\natexlab{}.
\newblock \showarticletitle{Note on the sampling error of the difference
  between correlated proportions or percentages}.
\newblock \bibinfo{journal}{\emph{Psychometrika}} \bibinfo{volume}{12},
  \bibinfo{number}{2} (\bibinfo{year}{1947}), \bibinfo{pages}{153--157}.
\newblock


\bibitem[OpenAI(2023)]%
        {openai2023gpt4}
\bibfield{author}{\bibinfo{person}{OpenAI}.} \bibinfo{year}{2023}\natexlab{}.
\newblock \bibinfo{title}{GPT-4 Technical Report}.
\newblock
\newblock
\showeprint[arxiv]{2303.08774}~[cs.CL]


\bibitem[Papineni et~al\mbox{.}(2002)]%
        {papineni2002bleu}
\bibfield{author}{\bibinfo{person}{Kishore Papineni}, \bibinfo{person}{Salim
  Roukos}, \bibinfo{person}{Todd Ward}, {and} \bibinfo{person}{Wei-Jing Zhu}.}
  \bibinfo{year}{2002}\natexlab{}.
\newblock \showarticletitle{Bleu: a method for automatic evaluation of machine
  translation}. In \bibinfo{booktitle}{\emph{Proceedings of the 40th annual
  meeting of the Association for Computational Linguistics}}.
  \bibinfo{pages}{311--318}.
\newblock


\bibitem[Raffel et~al\mbox{.}(2019)]%
        {raffel2019exploring}
\bibfield{author}{\bibinfo{person}{Colin Raffel}, \bibinfo{person}{Noam
  Shazeer}, \bibinfo{person}{Adam Roberts}, \bibinfo{person}{Katherine Lee},
  \bibinfo{person}{Sharan Narang}, \bibinfo{person}{Michael Matena},
  \bibinfo{person}{Yanqi Zhou}, \bibinfo{person}{Wei Li}, {and}
  \bibinfo{person}{Peter~J. Liu}.} \bibinfo{year}{2019}\natexlab{}.
\newblock \bibinfo{title}{Exploring the Limits of Transfer Learning with a
  Unified Text-to-Text Transformer}.
\newblock
\newblock
\showeprint[arxiv]{1910.10683}~[cs.LG]


\bibitem[Ren et~al\mbox{.}(2020)]%
        {Ren:codebleu}
\bibfield{author}{\bibinfo{person}{Shuo Ren}, \bibinfo{person}{Daya Guo},
  \bibinfo{person}{Shuai Lu}, \bibinfo{person}{Long Zhou},
  \bibinfo{person}{Shujie Liu}, \bibinfo{person}{Duyu Tang},
  \bibinfo{person}{Neel Sundaresan}, \bibinfo{person}{Ming Zhou},
  \bibinfo{person}{Ambrosio Blanco}, {and} \bibinfo{person}{Shuai Ma}.}
  \bibinfo{year}{2020}\natexlab{}.
\newblock \showarticletitle{CodeBLEU: a Method for Automatic Evaluation of Code
  Synthesis}.
\newblock \bibinfo{journal}{\emph{CoRR}}  \bibinfo{volume}{abs/2009.10297}
  (\bibinfo{year}{2020}).
\newblock
\urldef\tempurl%
\url{https://arxiv.org/abs/2009.10297}
\showURL{%
\tempurl}


\bibitem[Saroar and Nayebi(2023)]%
        {saroar2023developers}
\bibfield{author}{\bibinfo{person}{Sk~Golam Saroar} {and}
  \bibinfo{person}{Maleknaz Nayebi}.} \bibinfo{year}{2023}\natexlab{}.
\newblock \showarticletitle{Developers' Perception of GitHub Actions: A Survey
  Analysis}.
\newblock \bibinfo{journal}{\emph{arXiv preprint arXiv:2303.04084}}
  (\bibinfo{year}{2023}).
\newblock


\bibitem[Sutskever et~al\mbox{.}(2014)]%
        {sutskever2014sequence}
\bibfield{author}{\bibinfo{person}{Ilya Sutskever}, \bibinfo{person}{Oriol
  Vinyals}, {and} \bibinfo{person}{Quoc~V Le}.}
  \bibinfo{year}{2014}\natexlab{}.
\newblock \showarticletitle{Sequence to sequence learning with neural
  networks}.
\newblock \bibinfo{journal}{\emph{Advances in neural information processing
  systems}}  \bibinfo{volume}{27} (\bibinfo{year}{2014}).
\newblock


\bibitem[Svyatkovskiy et~al\mbox{.}(2020a)]%
        {svyatkovskiy2020intellicode}
\bibfield{author}{\bibinfo{person}{Alexey Svyatkovskiy},
  \bibinfo{person}{Shao~Kun Deng}, \bibinfo{person}{Shengyu Fu}, {and}
  \bibinfo{person}{Neel Sundaresan}.} \bibinfo{year}{2020}\natexlab{a}.
\newblock \showarticletitle{Intellicode compose: Code generation using
  transformer}. In \bibinfo{booktitle}{\emph{Proceedings of the 28th ACM Joint
  Meeting on European Software Engineering Conference and Symposium on the
  Foundations of Software Engineering}}. \bibinfo{pages}{1433--1443}.
\newblock


\bibitem[Svyatkovskiy et~al\mbox{.}(2020b)]%
        {svyatkovskiy2020fast}
\bibfield{author}{\bibinfo{person}{Alexey Svyatkovskiy},
  \bibinfo{person}{Sebastian Lee}, \bibinfo{person}{Anna Hadjitofi},
  \bibinfo{person}{Maik Riechert}, \bibinfo{person}{Juliana Franco}, {and}
  \bibinfo{person}{Miltiadis Allamanis}.} \bibinfo{year}{2020}\natexlab{b}.
\newblock \bibinfo{title}{Fast and Memory-Efficient Neural Code Completion}.
\newblock
\newblock
\showeprint[arxiv]{2004.13651}~[cs.SE]


\bibitem[Tufano et~al\mbox{.}(2022a)]%
        {tufano2022generating}
\bibfield{author}{\bibinfo{person}{Michele Tufano}, \bibinfo{person}{Dawn
  Drain}, \bibinfo{person}{Alexey Svyatkovskiy}, {and} \bibinfo{person}{Neel
  Sundaresan}.} \bibinfo{year}{2022}\natexlab{a}.
\newblock \showarticletitle{Generating accurate assert statements for unit test
  cases using pretrained transformers}. In
  \bibinfo{booktitle}{\emph{Proceedings of the 3rd ACM/IEEE International
  Conference on Automation of Software Test}}. \bibinfo{pages}{54--64}.
\newblock


\bibitem[Tufano et~al\mbox{.}(2019)]%
        {tufano2019empirical}
\bibfield{author}{\bibinfo{person}{Michele Tufano}, \bibinfo{person}{Cody
  Watson}, \bibinfo{person}{Gabriele Bavota}, \bibinfo{person}{Massimiliano~Di
  Penta}, \bibinfo{person}{Martin White}, {and} \bibinfo{person}{Denys
  Poshyvanyk}.} \bibinfo{year}{2019}\natexlab{}.
\newblock \showarticletitle{An empirical study on learning bug-fixing patches
  in the wild via neural machine translation}.
\newblock \bibinfo{journal}{\emph{ACM Transactions on Software Engineering and
  Methodology (TOSEM)}} \bibinfo{volume}{28}, \bibinfo{number}{4}
  (\bibinfo{year}{2019}), \bibinfo{pages}{1--29}.
\newblock


\bibitem[Tufano et~al\mbox{.}(2022b)]%
        {tufano2022using}
\bibfield{author}{\bibinfo{person}{Rosalia Tufano}, \bibinfo{person}{Simone
  Masiero}, \bibinfo{person}{Antonio Mastropaolo}, \bibinfo{person}{Luca
  Pascarella}, \bibinfo{person}{Denys Poshyvanyk}, {and}
  \bibinfo{person}{Gabriele Bavota}.} \bibinfo{year}{2022}\natexlab{b}.
\newblock \showarticletitle{Using Pre-Trained Models to Boost Code Review
  Automation}.
\newblock \bibinfo{journal}{\emph{arXiv preprint arXiv:2201.06850}}
  (\bibinfo{year}{2022}).
\newblock


\bibitem[Tufano et~al\mbox{.}(2021)]%
        {tufano2021towards}
\bibfield{author}{\bibinfo{person}{Rosalia Tufano}, \bibinfo{person}{Luca
  Pascarella}, \bibinfo{person}{Michele Tufano}, \bibinfo{person}{Denys
  Poshyvanyk}, {and} \bibinfo{person}{Gabriele Bavota}.}
  \bibinfo{year}{2021}\natexlab{}.
\newblock \showarticletitle{Towards automating code review activities}. In
  \bibinfo{booktitle}{\emph{2021 IEEE/ACM 43rd International Conference on
  Software Engineering (ICSE)}}. IEEE, \bibinfo{pages}{163--174}.
\newblock


\bibitem[Vasilescu et~al\mbox{.}(2015)]%
        {VasilescuFSE15}
\bibfield{author}{\bibinfo{person}{Bogdan Vasilescu}, \bibinfo{person}{Yue Yu},
  \bibinfo{person}{Huaimin Wang}, \bibinfo{person}{Premkumar~T. Devanbu}, {and}
  \bibinfo{person}{Vladimir Filkov}.} \bibinfo{year}{2015}\natexlab{}.
\newblock \showarticletitle{Quality and productivity outcomes relating to
  continuous integration in GitHub}. In
  \bibinfo{booktitle}{\emph{{ESEC/SIGSOFT} {FSE}}}. \bibinfo{publisher}{{ACM}},
  \bibinfo{pages}{805--816}.
\newblock


\bibitem[Vaswani et~al\mbox{.}(2017)]%
        {vaswani2017attention}
\bibfield{author}{\bibinfo{person}{Ashish Vaswani}, \bibinfo{person}{Noam
  Shazeer}, \bibinfo{person}{Niki Parmar}, \bibinfo{person}{Jakob Uszkoreit},
  \bibinfo{person}{Llion Jones}, \bibinfo{person}{Aidan~N Gomez},
  \bibinfo{person}{{\L}ukasz Kaiser}, {and} \bibinfo{person}{Illia
  Polosukhin}.} \bibinfo{year}{2017}\natexlab{}.
\newblock \showarticletitle{Attention is all you need}.
\newblock \bibinfo{journal}{\emph{Advances in neural information processing
  systems}}  \bibinfo{volume}{30} (\bibinfo{year}{2017}).
\newblock


\bibitem[Wang et~al\mbox{.}(2021b)]%
        {wang2021context}
\bibfield{author}{\bibinfo{person}{Haoye Wang}, \bibinfo{person}{Xin Xia},
  \bibinfo{person}{David Lo}, \bibinfo{person}{Qiang He},
  \bibinfo{person}{Xinyu Wang}, {and} \bibinfo{person}{John Grundy}.}
  \bibinfo{year}{2021}\natexlab{b}.
\newblock \showarticletitle{Context-aware retrieval-based deep commit message
  generation}.
\newblock \bibinfo{journal}{\emph{ACM Transactions on Software Engineering and
  Methodology (TOSEM)}} \bibinfo{volume}{30}, \bibinfo{number}{4}
  (\bibinfo{year}{2021}), \bibinfo{pages}{1--30}.
\newblock


\bibitem[Wang et~al\mbox{.}(2021a)]%
        {wang2021codet5}
\bibfield{author}{\bibinfo{person}{Yue Wang}, \bibinfo{person}{Weishi Wang},
  \bibinfo{person}{Shafiq Joty}, {and} \bibinfo{person}{Steven~CH Hoi}.}
  \bibinfo{year}{2021}\natexlab{a}.
\newblock \showarticletitle{Codet5: Identifier-aware unified pre-trained
  encoder-decoder models for code understanding and generation}.
\newblock \bibinfo{journal}{\emph{arXiv preprint arXiv:2109.00859}}
  (\bibinfo{year}{2021}).
\newblock


\bibitem[Watson et~al\mbox{.}(2020a)]%
        {WatsonTMBP20}
\bibfield{author}{\bibinfo{person}{Cody Watson}, \bibinfo{person}{Michele
  Tufano}, \bibinfo{person}{Kevin Moran}, \bibinfo{person}{Gabriele Bavota},
  {and} \bibinfo{person}{Denys Poshyvanyk}.} \bibinfo{year}{2020}\natexlab{a}.
\newblock \showarticletitle{On learning meaningful assert statements for unit
  test cases}. In \bibinfo{booktitle}{\emph{{ICSE} '20: 42nd International
  Conference on Software Engineering, Seoul, South Korea, 27 June - 19 July,
  2020}}. \bibinfo{publisher}{{ACM}}, \bibinfo{pages}{1398--1409}.
\newblock


\bibitem[Watson et~al\mbox{.}(2020b)]%
        {watson2020learning}
\bibfield{author}{\bibinfo{person}{Cody Watson}, \bibinfo{person}{Michele
  Tufano}, \bibinfo{person}{Kevin Moran}, \bibinfo{person}{Gabriele Bavota},
  {and} \bibinfo{person}{Denys Poshyvanyk}.} \bibinfo{year}{2020}\natexlab{b}.
\newblock \showarticletitle{On learning meaningful assert statements for unit
  test cases}. In \bibinfo{booktitle}{\emph{Proceedings of the ACM/IEEE 42nd
  International Conference on Software Engineering}}.
  \bibinfo{pages}{1398--1409}.
\newblock


\bibitem[Wilcoxon(1945)]%
        {wilcoxon}
\bibfield{author}{\bibinfo{person}{Frank Wilcoxon}.}
  \bibinfo{year}{1945}\natexlab{}.
\newblock \showarticletitle{Individual Comparisons by Ranking Methods}.
\newblock \bibinfo{journal}{\emph{Biometrics Bulletin}} \bibinfo{volume}{1},
  \bibinfo{number}{6} (\bibinfo{year}{1945}), \bibinfo{pages}{80--83}.
\newblock
\showISSN{00994987}


\bibitem[Zampetti et~al\mbox{.}(2022)]%
        {zampetti2022problems}
\bibfield{author}{\bibinfo{person}{Fiorella Zampetti},
  \bibinfo{person}{Vittoria Nardone}, {and} \bibinfo{person}{Massimiliano
  Di~Penta}.} \bibinfo{year}{2022}\natexlab{}.
\newblock \showarticletitle{Problems and solutions in applying continuous
  integration and delivery to 20 open-source cyber-physical systems}. In
  \bibinfo{booktitle}{\emph{Proceedings of the 19th International Conference on
  Mining Software Repositories}}. \bibinfo{pages}{646--657}.
\newblock


\bibitem[Zampetti et~al\mbox{.}(2020)]%
        {zampetti2020empirical}
\bibfield{author}{\bibinfo{person}{Fiorella Zampetti}, \bibinfo{person}{Carmine
  Vassallo}, \bibinfo{person}{Sebastiano Panichella}, \bibinfo{person}{Gerardo
  Canfora}, \bibinfo{person}{Harald Gall}, {and} \bibinfo{person}{Massimiliano
  Di~Penta}.} \bibinfo{year}{2020}\natexlab{}.
\newblock \showarticletitle{An empirical characterization of bad practices in
  continuous integration}.
\newblock \bibinfo{journal}{\emph{Empirical Software Engineering}}
  \bibinfo{volume}{25} (\bibinfo{year}{2020}), \bibinfo{pages}{1095--1135}.
\newblock


\end{thebibliography}

\end{document}